\documentclass[12pt]{article}

\usepackage[T1]{fontenc}
\usepackage[utf8]{inputenc}
\usepackage{ae,aecompl}
\usepackage{microtype}
\usepackage{csquotes}

\usepackage[margin=1in]{geometry}
\usepackage{amsmath}
\usepackage{amssymb}
\usepackage{amsfonts}
\usepackage{euscript}
\usepackage{amsthm}
\usepackage{mathtools}
\usepackage{bm}
\allowdisplaybreaks

\newtheorem{theorem}{Theorem}
\newtheorem{lemma}{Lemma}
\newtheorem{proposition}{Proposition}
\newtheorem{corollary}{Corollary}
\newtheorem{claim}{Claim}
\newtheorem{fact}{Fact}

\theoremstyle{definition}
\newtheorem{definition}{Definition}
\newtheorem{example}{Example}

\theoremstyle{remark}
\newtheorem{remark}{Remark}

\usepackage{graphicx}
\usepackage{array}
\usepackage{booktabs}
\usepackage{enumitem}
\usepackage{rotating}
\usepackage{colortbl}
\usepackage{tikz}
\usetikzlibrary{patterns}
\usepackage[title]{appendix}
\graphicspath{{figures/}}

\usepackage{color}
\definecolor{webgreen}{rgb}{0,0.4,0}
\definecolor{webbrown}{rgb}{0.6,0,0}
\definecolor{darkblue}{rgb}{0,0,0.7}
\definecolor{darkred}{rgb}{0.7,0,0}
\definecolor{darkgreen}{rgb}{0,0.7,0}

\usepackage[round,authoryear]{natbib}
\usepackage[pdfborder={0 0 0}]{hyperref}
\hypersetup{colorlinks,citecolor=darkred,filecolor=black,linkcolor=darkblue,
            urlcolor=webgreen,pdfstartview=FitH,
            pdfauthor={Karan Maniyar},pdftitle={Fair Combinatorial Auctions}}

\DeclareMathOperator{\Core}{Core}
\newcommand{\R}{\mathbb{R}}

\newcommand{\1}{\mathbf{1}}

\title{\vspace{-1.5cm}{\bf Fair Combinatorial Auctions}}
\author{Karan Maniyar\thanks{Indian Statistical Institute, Delhi, \texttt{karan25q@isid.ac.in}.\newline
I am indebted to Debasis Mishra for his guidance on this paper. I am especially grateful to Arunava Sen, Jay Sethuraman, Sushil Bikhchandani, Salil Sharma, Vilok Taori and Sarvesh Choushetti for insightful comments that have greatly improved the paper. I also thank participants of the Theory Reading Group at ISI Delhi for helpful discussions, and Refine (\texttt{refine.ink}) for comments and corrections on an earlier draft. }}
\date{26 November 2026}
\begin{document}
\hypersetup{pageanchor=false}\emergencystretch=1em
\maketitle
\begin{abstract}
The Vickrey--Clarke--Groves (VCG) mechanism can leave the seller with no revenue, even when its outcome is stable.
We ask how the gains from a combinatorial auction can be divided as fairly as possible between the seller and the buyers, among stable (core) outcomes.
We show that, when buyers are substitutes, the core contains a unique Lorenz-dominant vector: every player is paid a common amount, except that no buyer receives more than her marginal contribution.
The seller thus receives at least an equal share of the surplus.
Since this division generally cannot be reached by a strategy-proof auction, we construct a complete-information bargaining game in which every subgame-perfect equilibrium yields the Lorenz-dominant vector.
\end{abstract}
\noindent\textsc{Keywords:} Combinatorial auctions; Lorenz dominance; core; fairness; buyers are substitutes; non-cooperative implementation.

\thispagestyle{empty}
\clearpage\pagenumbering{arabic}\hypersetup{pageanchor=true}
\section{Introduction}\label{sec:introduction}
In 2016--17, the US Federal Communications Commission (FCC) ran the Broadcast Incentive Auction to move spectrum from television broadcasters to mobile carriers. After a first stage fixed how many frequency blocks each bidder won, an assignment stage decided which specific blocks each winner received and priced them by the Vickrey--Clarke--Groves (VCG) mechanism. Table~\ref{tab:incentive-auction} reports this stage in the area covering DeKalb, Illinois and Kenosha, Wisconsin.

\begin{table}[!htb]
\centering
\caption{VCG assignment stage, DeKalb, IL and Kenosha, WI}\label{tab:incentive-auction}
\smallskip
\begin{tabular}{lcrr}
\toprule
Bidder & Blocks received & Bid for these blocks (US\$) & Payment (US\$)\\
\midrule
T-Mobile        & B, C, D & 3,500,000 & 0\\
U.S.\ Cellular  & E, F, G & 8,700,000 & 0\\
Dish Network    & A       & 0         & 0\\
\midrule
Total           &         & 12,200,000 & 0\\
\bottomrule
\end{tabular}

\smallskip
{\footnotesize Dish bid about US\$237,000 for block G but received block A. Source: \citet{ausubel2026vcg}.\par}
\end{table}

\noindent T-Mobile and U.S.\ Cellular together bid US\$12.2 million for the blocks they won, and neither paid anything. By the textbook standard, nothing went wrong: the blocks went to the bidders who valued them most, and no bidder could gain by bidding other than her true value. Yet the outcome raises two problems. The first concerns the division of the gains: the buyers kept the entire surplus, and the seller, here the US government, received none of it. The second is that the outcome invites a side deal. Dish bid about US\$237,000 for block G, which went to U.S.\ Cellular for free. Selling block G to Dish at any price below that bid would have made both the seller and Dish better off. \citet{ausubel2026vcg} find three such zero-revenue areas in this auction. This paper asks: keeping the goods with those who value them most, how should the gains be divided between the seller and the buyers, so that the division is as fair as possible and no group of players would rather strike its own deal? We show that, when buyers are substitutes, the fairest stable division is unique and simple: every player is paid a common amount, except that no buyer receives more than her marginal contribution. The seller always receives at least an equal share of the surplus. We also construct a bargaining game whose every equilibrium yields this division.

Why would a government sell its goods this way? A planner wants the goods to go where they create the most value, and since only the buyers know their values, she also needs them to bid truthfully and take part willingly. VCG delivers all of this: it is efficient, truthful bidding is a dominant strategy (DSIC), no buyer loses by taking part (ex post individual rationality), and no buyer is paid to show up (nonnegative payments). On a regular domain of valuations, it is the \emph{only} mechanism with these four properties.\footnote{A domain of valuation profiles is \emph{regular} if it is a product of convex sets, one for each buyer, each containing the zero valuation, which values every bundle at zero. The set of all normalised, monotone valuations is an example. On a regular domain, every efficient, DSIC mechanism uses Groves payments \citep{holmstrom1979groves}; ex post individual rationality and nonnegative payments then pin these down as VCG payments (Section~\ref{sec:implementation}).} The zero revenue in Table~\ref{tab:incentive-auction} is therefore not a flaw in how the FCC ran the auction but what these properties deliver. A fairer division of the surplus requires giving something up.

Where should the planner look for a fair division? We look among the stable outcomes. An outcome is stable if no group of players, the seller together with some buyers, can do better by leaving the auction and trading among themselves, as the seller and Dish could have done. Stability is our main requirement: a fair division is of little use if a group of players can overturn it with a side deal. The stable outcomes form the core of the cooperative game induced by the auction. Every stable outcome is efficient, individually rational and has nonnegative payments, so stability keeps three of VCG's four properties. VCG itself, however, need not be stable. \citet{rothkopf1990why} were among the first to point out this fragility, and \citet{ausubel2006lovely} call the Vickrey auction ``lovely but lonely''. \citet{bikhchandani2002package} show that the VCG outcome is stable if and only if buyers are substitutes. Even then, stability secures the seller little: in the incentive auction, ruling out the deal with Dish takes payments of only about US\$237,000 \citep{ausubel2026vcg}. The planner also cannot choose a different stable division while keeping truthful bidding: on a regular domain, every efficient, DSIC mechanism whose outcome is stable at every valuation profile must charge VCG payments.

A stable division other than VCG's therefore requires giving up DSIC. This is the only property the planner can give up, since stability itself guarantees the other three. Doing so opens a typically large set of stable outcomes, of which VCG is one extreme: it gives every buyer her largest possible payoff and the seller her smallest \citep{ausubel2002ascending}. Take three identical licences and two buyers who each want one, with values 10 and 8. VCG charges nothing, so the payoffs to (seller, buyer 1, buyer 2) are $(0,10,8)$. This is stable, but so is $(6,6,6)$. Stability rules out side deals but leaves the division of the surplus open, so we ask which stable outcome is the fairest. We treat the seller as a full player in this comparison, not as a leftover claimant. When the seller is a government, low revenue is not a private transfer between two parties. It is a loss to the public exchequer, as the incentive auction illustrates.

Fairness has many definitions: the Rawlsian max--min criterion, Nash welfare, and requirements such as anonymity. We use the strongest one, which reconciles all of these: Lorenz dominance. One payoff vector Lorenz dominates another with the same total if, for every $k$, its $k$ poorest players together receive at least as much. For example, $(40,40,20)$ Lorenz dominates $(50,30,20)$: the poorest player gets 20 in both, but the two poorest get 60 instead of 50. By a classical result of \citet{hardy1929some}, for payoff vectors $x$ and $y$ with the same total,
\[
x\succeq_L y\quad\Longleftrightarrow\quad
\sum_{i\in N}U(x_i)\geq\sum_{i\in N}U(y_i)\quad\text{for every increasing, concave }U
\]
where $U$ is the planner's welfare function. A Lorenz-dominant vector therefore reconciles these criteria in a single payoff vector.

The catch is that Lorenz dominance is only a partial order, and most sets contain no such vector. The vectors $(7,1,2)$ and $(5,5,0)$ have the same total, yet neither dominates the other. One positive result is known. \citet{dutta1989concept} show that the core of every convex game, in which a player adds weakly more to a larger coalition, contains a Lorenz-dominant division. Auction games, however, are not convex in general (Remark~\ref{rem:box} gives an example), so this result does not apply directly. Existence of a Lorenz-dominant vector in the core of an auction is therefore far from obvious.

\paragraph{First contribution.} Formally, if buyers are substitutes, the same condition that makes VCG stable, the core contains a unique Lorenz-dominant vector. Let $m_i$ be buyer $i$'s marginal contribution to the total surplus $W(N)$, which is exactly her VCG payoff. There is a single number $\sigma$, the unique solution of
\[
x+\sum_{i\in B}\min\{x,m_i\}=W(N),
\]
such that the seller receives $\sigma$ and each buyer $i$ receives $\min\{\sigma,m_i\}$ (Theorem~\ref{thm:ldv-existence}). When buyers are substitutes, each buyer's marginal contribution caps her payoff, and these individual caps alone determine the core. The most equal stable division therefore raises all payoffs to a common level, stopping each buyer at her cap. Equivalently, it is a small change to VCG: shave the payoffs of the best-off buyers and hand the shavings to the seller, until she stands level with the tallest remaining buyer. In the licence example, $\sigma=6$ and the outcome is $(6,6,6)$, so the seller's payoff rises from 0 to 6. In general, the seller receives at least $W(N)/(|B|+1)$ and never less than any buyer, whereas under VCG she can receive zero. The substitutes condition is sufficient but not necessary: without it, a Lorenz-dominant vector may or may not exist.

\paragraph{When is VCG already fair?} The Lorenz-dominant vector equals the VCG payoff vector if and only if the seller's VCG payoff is at least every buyer's marginal contribution (Corollary~\ref{cor:vcg-ldv}). VCG is then efficient, DSIC, stable and fair at once.

\noindent\textbf{Second contribution.} When the Lorenz-dominant vector differs from the VCG payoff vector, no efficient, DSIC auction implements it. We therefore turn to complete information and ask whether the fair division can emerge from bargaining rather than be imposed by the planner. In our game, the goods are allocated efficiently and only payments are bargained over: the seller proposes a division of the surplus and buyers respond in turn; a buyer who rejects leaves with an exit payment, and bargaining continues among the rest. The rules never mention the target vector, let the seller propose any division, and never destroy surplus. Yet every subgame-perfect equilibrium yields the Lorenz-dominant vector (Theorem~\ref{thm:seller-proposer-implementation}). The seller holds all the power to propose, but the exit payments, built only from marginal contributions and her own claim, stop her from taking more than her fair share. Unlike earlier bargaining foundations for egalitarian divisions \citep{chatterjee1993noncooperative}, the result needs no discounting or stationarity.

\paragraph{Organisation.} Section~\ref{sec:literature} discusses related literature. Section~\ref{sec:model} sets up the combinatorial auction, the VCG mechanism and the cooperative game it induces. Section~\ref{sec:fair-auctions} introduces Lorenz dominance, proves the existence and characterisation of the Lorenz-dominant vector, and discusses when it coincides with VCG, whether the substitutes condition is necessary, and which valuations guarantee it. Section~\ref{sec:implementation} explains why truthful implementation is impossible in general and presents the seller-proposer game and its implementation result. Appendix~\ref{app:existence} proves Theorem~\ref{thm:ldv-existence}, Appendix~\ref{app:implementation} completes the proof of Theorem~\ref{thm:seller-proposer-implementation}, Appendix~\ref{app:pooling} gives a sequential pooling algorithm that computes the vector, and Appendix~\ref{app:examples} contains the examples on the necessity of the substitutes condition.

\section{Related literature}\label{sec:literature}
\noindent\textbf{Stability and payments in combinatorial auctions.} \citet{day2008core} study auctions that select core outcomes with respect to reported values, addressing instability and low revenues when the VCG outcome is outside the core. Their minimum-revenue approach keeps payments as low as the core permits to limit incentives to misreport. \citet{day2012quadratic} refine this selection by choosing, among minimum-revenue core payments, the vector nearest to VCG payments in Euclidean distance. These rules retain VCG whenever it lies in the reported core. Our fairness question remains even when VCG is stable: its division of surplus can still leave the seller with very little. We include the seller in the fairness comparison and seek the most equal division in the core of the game defined by true valuations.

\noindent\textbf{Existence of Lorenz-dominant vectors.} \citet{dutta1989concept} establish existence in the core of convex cooperative games. \citet{mishra2025teacher} find an integral Lorenz-dominant post-transfer deficit vector in a teacher-transfer problem. Their network formulation also induces a convex game; the additional difficulty is to respect the indivisibility of teachers. In matching, \citet{bogomolnaia2004random} obtain a Lorenz-dominant vector of probabilities of acceptable matches under dichotomous preferences. Although the object of comparison is a feasible probability vector, their construction also uses a convex-game representation. In bankruptcy problems, \citet{bosmans2011lorenz} compare divisions of an estate subject to individual claims. They show that constrained equal awards Lorenz dominates every claims rule that preserves the ordering of awards by claims. Our setting connects to both strands. The auction game need not be convex. When buyers are substitutes, however, its core is the set of efficient payoff vectors in which no buyer receives more than her marginal contribution. It therefore coincides with the core of an auxiliary convex game, and existence of a Lorenz-dominant vector can alternatively be derived from \citet{dutta1989concept} (Appendix~\ref{app:existence-convex}). On this set, the Lorenz-dominant vector is the constrained equal awards division of $W(N)$, with each buyer's marginal contribution as her claim. The contribution of Theorem~\ref{thm:ldv-existence} is to identify this structure in combinatorial auctions, to tie it to the condition under which VCG is stable, and to show that the seller always receives at least an equal share.

\noindent\textbf{Non-cooperative implementation.} The Nash program, initiated by \citet{nash1953two} and surveyed by \citet{serrano2005fifty}, seeks non-cooperative foundations for cooperative solution concepts. For strictly convex games, \citet{chatterjee1993noncooperative} show that efficient stationary equilibrium payoffs converge to the Dutta--Ray egalitarian allocation as discounting vanishes. Other work implements different divisions of the cooperative surplus. \citet{perez2001bidding} implement the Shapley value for zero-monotonic games through bidding for the right to propose, followed by offers and sequential responses. \citet{ju2009implementing} generalise this approach to implement the Shapley, consensus and equal surplus values by varying the proposer's bargaining rights. Our game keeps the seller as proposer and gives rejecting buyers exit payments. It implements the Lorenz-dominant vector exactly in every subgame-perfect equilibrium, without discounting or a restriction to stationary strategies.

\section{The combinatorial auction model}
\label{sec:model}

We consider a combinatorial auction model with a single seller and a finite set of buyers $B=\{1,\ldots,n\}$. The seller owns a finite set of indivisible goods $G$. Her value for retaining the goods is normalised to zero. Let $N=\{s\}\cup B$
denote the set of all players, where $s$ denotes the seller.

\subsection{Buyer valuations and efficient allocation}

Each buyer $i \in B$ has a valuation $v_i:2^G\longrightarrow \mathbb{R}_{+}$, where $v_i(A)$ is buyer $i$'s value from receiving the bundle $A\subseteq G$. We assume that the values are normalised to zero for empty bundle and monotone : 
$$v_i(\varnothing)=0
\qquad\text{and}\qquad
v_i(A)\leq v_i(A')
\quad\text{whenever }A\subseteq A'$$

Buyers have quasilinear preferences: if buyer $i$ receives bundle $A_i$ and makes a payment $p_i$, her payoff is $v_i(A)-p_i$. A \emph{feasible} allocation $X = (X_s, X_1, \dots, X_n)$ is a partition of the set of goods $G$,
\[
\bigcup_{i \in \{s, 1, \dots, n\}} X_i = G \quad \text{and} \quad X_i \cap X_j = \emptyset \quad \forall i \neq j.
\]

Note that the seller need not allocate all the goods and hence $X_s$ denotes the set of goods still remaining with the seller. Let $\mathcal{P}$ denote the set of all feasible allocations of $G$. For any subset $S\subseteq B$ of buyers, let $\mathcal{P}(S)$ denote the set of feasible allocations in which only buyers in $S$ can receive nonempty bundles, 
$
\mathcal{P}(S) = \left\{ X \in \mathcal{P} \ \middle|\ X_j = \emptyset \quad \text{for every }j\in B\setminus S \right\}
$. Define the maximum surplus that can be generated using those buyers by,
\[
U(S) = \max_{X \in \mathcal{P}(S)} \sum_{i \in S} v_i(X_i)
\]

with $U(\varnothing)=0$. Since the set of goods is finite, the maximum exists. The total surplus available in the auction is $U(B)$. An allocation \(X^*\) is said to be \emph{efficient} if
\[
\sum_{i\in B}v_i(X_i^*)=U(B)
\]

\subsection{VCG Mechanism}
\label{subsec:2.2}

It is well known that the Vickrey-Clarke-Groves (VCG) mechanism implements the efficient allocation rule in dominant strategies. In a combinatorial auction, the mechanism asks each buyer $i\in B$ to report her valuation
$v_i:2^G\longrightarrow\mathbb{R}_{+}$. On the basis of these reports, it selects an efficient allocation. The payment charged to buyer $i$ is the externality that her presence imposes on the other buyers, given by
\[
p_i^{VCG}
=
U(B\setminus\{i\})
-
\sum_{j\neq i}v_j(X_j^*)
\]

Consequently, buyer $i$'s payoff is
\begin{align*}
\pi_i^{VCG}
&=
v_i(X_i^*)-p_i^{VCG} 
=
U(B)-U(B\setminus\{i\}).
\end{align*}
The expression on the right hand side is the marginal contribution of buyer $i$. Let $m_i$ denote the marginal contribution of buyer $i$, which is increase in the gains from trade due to $i$,
\[
m_i \equiv U(B)-U(B\setminus\{i\})
\]
Thus, every buyer receives her marginal contribution to the total surplus, $\pi_i^{VCG}=m_i$.
Seller receives payoff equal to the sum of the payments of the buyers,

\begin{align*}
    \pi_s^{VCG} &= \sum_{i \in B} p_i^{VCG} 
                = \sum_{i \in B} U(B \setminus \{i\}) - (n-1) U(B) 
                = U(B)-\sum_{i\in B}m_i
\end{align*}

Hence the seller's payoff can also be interpreted as residual surplus out of $U(B)$ after every buyer is paid her marginal contribution $m_i$.
The resulting Vickrey payoff vector is therefore
\[
\pi^{VCG}
=
\left(
U(B)-\sum_{i\in B}m_i,\,
(m_i)_{i\in B}
\right)
\]

The VCG mechanism is efficient, dominant-strategy incentive compatible, and ex post individually rational, which makes the VCG mechanism a natural benchmark for the design of combinatorial auctions. However, the following example from \citet{ausubel2006lovely} shows that the Vickrey payoff vector need not be coalitionally stable.

\begin{example}
\label{ex:vcg-core-failure}
Let $G = \{a,b\}$ and $B = \{1,2,3\}$. Let the valuations be,

\[
 v_1(\{a\}) = 0,\quad  v_1(\{b\}) = 0,\quad v_1(\{a,b\}) = 2
\]
\[
v_2(\{a\}) = 2,\quad v_2(\{b\}) = 0,\quad v_2(\{a,b\}) = 2
\]
\[
v_3(\{a\}) = 0,\quad v_3(\{b\}) = 2,\quad v_3(\{a,b\}) = 2
\]
Buyer $1$ values only the full bundle, while buyers $2$ and $3$ each value a distinct good. The efficient allocation assigns $a$ to buyer $2$ and $b$ to buyer $3$, generating a total surplus of $4$. The Vickrey payoff vector is $\pi^{VCG}=(0,0,2,2)$. However, the seller and buyer $1$ can generate a surplus of $2$ by allocating both goods to buyer $1$, whereas
\[
\pi_s^{VCG}+\pi_1^{VCG}=0<U(\{1\})=2.
\]
They can therefore divide this surplus so that both are strictly better off.
\end{example}
In the next section, we formalise coalitional stability through the core of the transferable utility cooperative game induced by the auction.

\subsection{Cooperative game induced by the combinatorial auction}

The combinatorial auction induces a cooperative game \((N,W)\), where $N=\{s\}\cup B$ is the set of players and $W$ is the coalitional value function. For every coalition \(C\subseteq N\), 
\[
W(C)
=
\begin{cases}
U(C\cap B), & \text{if }s\in C,\\[3pt]
0,          & \text{if }s\notin C.
\end{cases}
\]

In other words, $W(C)$ is the maximum value that the coalition $C$ can generate on their own, which is positive only if it contains the seller, because the seller initially owns all the goods. A payoff vector is written as $x=(x_s,(x_i)_{i\in B})$.
It is said to be an imputation if it distributes the entire surplus and gives every player at least her stand alone payoff,
\[
x_s+\sum_{i\in B}x_i=W(N)
\qquad\text{and}\qquad
x_j\geq 0
\quad\text{for every }j\in N.
\]
The core of the auction game is,
\[
\Core(N,W)
=
\left\{
x\in\mathbb{R}^{N}:
\sum_{j\in N}x_j=W(N)
\ \text{and}\
\sum_{j\in C}x_j\geq W(C)
\ \text{for every }C\subseteq N
\right\}.
\]

The inequality associated with a coalition \(C\) requires that its members collectively receive at least as much as they could generate by leaving the grand coalition and acting on their own. Note that the core of this game is always nonempty as the payoff vector $x =(x_s,x_1,\ldots,x_n)$, where 
\[
x_s=W(N)
\qquad\text{and}\qquad
x_i=0
\quad\text{for every }i\in B
\]

always lies in the core. 
Thus, non-emptiness of core is not an issue in the game induced by the combinatorial auction with a single seller. Now we turn to the question: \enquote{When does Vickery payoff vector $\pi^{VCG}$ belong to the core?}. This question has been answered in \cite{ausubel2002ascending,bikhchandani2002package}.

First note that by definition of $W$,  $m_i$ as defined in the previous subsection also denotes the marginal contribution of buyer $i$ in the coalitional game $W$. That is, 

\[
m_i \equiv U(B)-U(B\setminus\{i\}) = W(N)-W(N\setminus\{i\})
\]

Hence, the Vickrey payoff vector $\pi^{VCG}$ can be rewritten in terms of the coalitional value function $W$ as,

\[
\pi^{VCG}
=
\left(
W(N)-\sum_{i\in B}m_i,\,
(m_i)_{i\in B}
\right)
\]

Let $m_T$ be the marginal contribution of the set of buyers $T\subseteq B$, 

\[
m_T \equiv  W(N)-W(N\setminus T)
\]

\begin{definition}\label{def:substitutes}
    Buyers are said to be \emph{substitutes} if for all $T\subseteq B$, 

\[
m_T \geq  \sum_{i \in T} m_i
\]
    
\end{definition}

All marginal contributions in this condition are evaluated at the grand coalition $N$. The loss from removing a group of buyers must be at least the sum of the losses from removing its members individually.

Buyer submodularity is strictly stronger: it is sufficient for buyers to be substitutes, but is not necessary.\footnote{The coalitional value function $W$ is \emph{buyer submodular} if, for all sets of buyers $S\subseteq S'\subseteq B$ and every buyer $i\in B\setminus S'$, $W(\{s\}\cup S\cup\{i\})-W(\{s\}\cup S)\geq W(\{s\}\cup S'\cup\{i\})-W(\{s\}\cup S')$. Equivalently, $U$ is submodular on $B$: a buyer adds weakly less surplus to a larger group of buyers. The buyers-are-substitutes condition requires a comparable inequality only for groups of buyers removed from the grand coalition.}

\citet{bikhchandani2002package} show that buyers are substitutes if and only if the VCG payoff belongs to the core:
\begin{equation}\label{eq:substitutes-vcg-core}
\text{buyers are substitutes}\quad\Longleftrightarrow\quad\pi^{VCG}\in\Core(N,W).
\end{equation}
In particular, when buyers are substitutes, the Vickrey payoff vector lies in the core.

\section{Fair Combinatorial Auctions}\label{sec:fair-auctions}
When buyers are substitutes, the Vickrey payoff vector belongs to the core, but it gives every buyer her largest core payoff and the seller her smallest \citep{ausubel2002ascending}. Stability therefore leaves a distributional question: which core payoff provides the fairest division of the surplus? The incentive auction discussed in the introduction illustrates why the seller's payoff matters in this comparison. We compare divisions of the fixed surplus $W(N)$ among all players, including the seller, using Lorenz dominance.

Let \(\pi\in\mathbb{R}^{|N|}_{+}\) be a payoff vector, and let
\[
    \pi_{(1)}\leq\pi_{(2)}\leq\cdots\leq\pi_{(|N|)}
\]
denote its components arranged from the smallest to the largest.

\begin{definition}
\label{def:lorenz-dominance}
Let \(\pi,\pi'\in\mathbb{R}^{|N|}_{+}\) satisfy
\[
    \sum_{j\in N}\pi_j=\sum_{j\in N}\pi'_j.
\]
We say that \(\pi\) \emph{Lorenz dominates} \(\pi'\), denoted by
\(\pi\succeq_L\pi'\), if
\[
    \sum_{r=1}^{k}\pi_{(r)}
    \geq
    \sum_{r=1}^{k}\pi'_{(r)}
    \qquad
    \text{for every }k=1,\ldots,|N|.
\]
\end{definition}

Thus, Lorenz dominance requires the $k$ least-paid players to receive at least as much collectively, for every $k$. The following fact shows that Lorenz dominance does not depend on one particular way of measuring inequality.
\begin{fact}[\cite{hardy1929some}]\label{fact:hardy-littlewood-polya}
Let $\pi,\pi'\in\mathbb R_+^{|N|}$ have the same total payoff. Then $\pi\succeq_L\pi'$ if and only if
\[
\sum_{j\in N}U(\pi_j)\geq\sum_{j\in N}U(\pi'_j)
\]
for every continuous, increasing, and concave function $U:[0,W(N)]\to\mathbb R$.
\end{fact}
On the nonempty, compact core, any continuous welfare criterion attains a maximum. Different criteria, however, may select different payoff vectors. Fact~\ref{fact:hardy-littlewood-polya} identifies a stronger requirement: a vector that Lorenz dominates every other core payoff simultaneously maximises $\sum_{j\in N}U(\pi_j)$ for every continuous, increasing, and concave function $U$. Such a vector reconciles this entire class of welfare criteria, allowing the choice of a fair division without selecting one particular criterion. The class contains the familiar criteria. The utilitarian criterion $U(x)=x$ is indifferent among all core payoffs, since they share the same total. Nash welfare corresponds to $U(x)=\log x$, and the constant-elasticity criteria to $U(x)=x^{1-\alpha}/(1-\alpha)$ for $\alpha>0$, $\alpha\neq1$.\footnote{For $\log x$ and for $\alpha>1$, $U$ is not finite at $0$. Lorenz dominance still implies a weakly larger sum for these criteria, by applying Fact~\ref{fact:hardy-littlewood-polya} to $U(x+\varepsilon)$ and letting $\varepsilon\to0$.} The Rawlsian max--min criterion is captured directly: the case $k=1$ of Definition~\ref{def:lorenz-dominance} requires the poorest player under $\pi$ to receive at least as much as the poorest player under $\pi'$.
\begin{definition}\label{def:lorenz-dominant-core}
A payoff vector $\pi^L\in\Core(N,W)$ is called a \emph{Lorenz-dominant vector} (LDV) if
\[\pi^L\succeq_L\pi\qquad\text{for every }\pi\in\Core(N,W).\]
\end{definition}
The existence of an LDV is not guaranteed by compactness. Lorenz dominance is only a partial order: two core payoffs may be incomparable, with neither providing a uniformly more equal division. It is therefore necessary to distinguish a payoff that dominates every other core payoff from one that is merely undominated.
\begin{definition}
A core payoff is \emph{Lorenz maximal} if it is not strictly Lorenz dominated by another core payoff.
\end{definition}
Lorenz-maximal payoffs exist in the nonempty, compact core, but need not dominate all other core payoffs. Thus, existence of a maximiser for each welfare criterion separately does not ensure the existence of a common maximiser for all of them. A LDV meets precisely this stronger requirement.

\citet{dutta1989concept} show that convex games admit a core allocation that Lorenz dominates every other core allocation. Our first main result establishes existence and uniqueness in auction games where buyers are substitutes. Such games need not be convex (Remark~\ref{rem:box}). The condition that ensures the stability of the Vickrey payoff therefore also guarantees that the core contains a division reconciling the entire class of welfare criteria in Fact~\ref{fact:hardy-littlewood-polya}.

\subsection{The Main Result}
\begin{theorem}\label{thm:ldv-existence}
Suppose buyers are substitutes. Then $\Core(N,W)$ contains a unique Lorenz-dominant vector $\pi^{LDV}$, given by
\[\pi_s^{LDV}=\sigma,\qquad\pi_i^{LDV}=\min\{\sigma,m_i\}\quad\text{for every }i\in B.\]
Moreover, the seller's payoff $\sigma$ has two equivalent characterisations:
\begin{enumerate}[label=(\roman*)]
\item $\sigma$ is the unique nonnegative solution to
\[x+\sum_{i\in B}\min\{x,m_i\}=W(N).\]
\item $\sigma$ is given by
\[\sigma=\max_{T\subseteq B}\frac{W(N)-\sum_{i\in T}m_i}{|B\setminus T|+1}.\]
\end{enumerate}
\end{theorem}

The theorem tells us that the Lorenz-dominant vector has a simple structure. A
single number $\sigma$ determines the entire vector: a buyer with $m_i<\sigma$
receives $m_i$, and every other player, including the seller, receives
$\sigma$. Thus the surplus is divided at a common level, and a buyer is paid
below that level only when her marginal contribution does not permit more. It
is useful to compare this vector with the Vickrey payoff vector
\[
\pi^{VCG} = \Big(W(N) - \sum_{i \in B} m_i,\ (m_i)_{i \in B}\Big),
\]
under which every buyer receives her full marginal contribution and the seller
receives the remainder. The two vectors agree on the buyers with
$m_i<\sigma$. They differ on the buyers with $m_i\geq\sigma$: each such buyer
receives $\sigma$ rather than $m_i$, and the difference $m_i-\sigma$ is
transferred to the seller. The correction is therefore confined to the buyers
whose marginal contribution exceeds the common level. Note also that it does not
change the sale of the goods. Both vectors divide the same total surplus
$W(N)$, and both lie in the core, so the more balanced division is not obtained
at the cost of efficiency or stability.

Return to the example of three identical licences and two unit-demand buyers
with values $10$ and $8$ discussed above. Here $m_1=10$, $m_2=8$ and
$W(N)=18$, so $\sigma$ solves $x+\min\{x,10\}+\min\{x,8\}=18$, which gives
$\sigma=6$. The Lorenz-dominant vector is therefore $(6,6,6)$. The seller
receives $6$ rather than the $0$ assigned by VCG, and the balanced vector
mentioned earlier is exactly the Lorenz-dominant one. This improvement is not
specific to the example. Since $x+\sum_{i\in B}\min\{x,m_i\}\leq(|B|+1)x$ for
every $x$, the level $\sigma$ satisfies
\[
\sigma \geq \frac{W(N)}{|B|+1}.
\]
Thus the seller receives at least an equal share of the surplus, whatever the
marginal contributions happen to be. Under VCG her payoff is the residual
$W(N)-\sum_{i\in B}m_i$, which can be zero, as both the example above and the incentive auction illustrate. Note also that $\pi^{LDV}_s=\sigma\geq
\pi^{LDV}_i$ for every buyer $i$, so the seller is never paid less than a
buyer. This is not imposed. It follows from the fact that $W(C)=0$ for every
coalition $C$ not containing the seller.

The second characterisation of the seller's payoff $\sigma$ is in the spirit of the levelling algorithm of \citet{dutta1989concept}, which finds the Lorenz-dominant vector in convex games. Fix any group of buyers $T$, pay each of them her full marginal contribution $m_i$, and split what remains equally between the seller and the buyers outside $T$. Each of these remaining players receives
\[\frac{W(N)-\sum_{i\in T}m_i}{|B\setminus T|+1}.\]
The theorem says that no group consisting of the seller and a subset of buyers can obtain an average payoff above $\sigma$ after paying all remaining buyers their full marginal contributions. This bound is attained by the seller together with the buyers whose marginal contributions are at least $\sigma$, precisely the richest group in the Lorenz-dominant vector.

\begin{remark}[Why the substitutes condition suffices]\label{rem:box}
Every core payoff satisfies $x_j\geq0$ for every player and $x_i\leq m_i$ for every buyer (Lemma~\ref{lem:cap}). When buyers are substitutes, these individual bounds are also sufficient: every coalitional constraint is implied by them (Lemma~\ref{lem:box}). The core is then a box cut by the efficiency hyperplane,
\begin{equation}\label{eq:core-box}
\Core(N,W)=\Big\{x\in\R^N_+:\sum_{j\in N}x_j=W(N)\text{ and }x_i\leq m_i\text{ for every }i\in B\Big\}.
\end{equation}
Consequently, the core of $(N,W)$ coincides with the core of a convex game: the game $W^V$ that assigns to every coalition containing the seller the total VCG payoff of its members. The auction game itself need not be convex. In the auction of Example~\ref{ex:two-buyer-seller-game}, $W(\{s,1\})+W(\{s,2\})=80+40=120>100=W(N)+W(\{s\})$. Theorem~\ref{thm:ldv-existence} could therefore also be derived from \citet{dutta1989concept} applied to $W^V$. Appendix~\ref{app:existence-convex} gives this argument.

The reduction shows exactly what the substitutes condition does: it is the condition under which the individual caps alone describe the core. When it fails, the box in \eqref{eq:core-box} can be strictly larger than the core, and its most equal point need not be stable. In Example~\ref{ex:vcg-core-failure}, the Lorenz-dominant point of the box is $(4/3,0,4/3,4/3)$, which gives coalition $\{s,1\}$ only $4/3<2=W(\{s,1\})$. Note also that on the box, $\pi^{LDV}$ is the constrained equal awards division of $W(N)$ in which buyer $i$ claims $m_i$ and the seller claims $W(N)$ \citep{bosmans2011lorenz}. The seller's claim never binds, since $\sigma\leq W(N)$.
\end{remark}

\begin{remark}\label{rem:pooling}
The vector $\pi^{LDV}$ can also be obtained through a simple equalization procedure. Start from the Vickrey payoff, where every buyer $i$ receives her marginal contribution $m_i$, the largest payoff she can receive in the core (by Lemma~\ref{lem:cap}). Whenever the seller is poorer than the richest buyers, transfer payoff from those buyers to the seller until either they are pooled with the seller or their payoff reaches that of the next-richest buyers, in which case the latter join the pool and the procedure continues. Each such transfer is progressive and therefore moves the payoff vector upward in the Lorenz ordering (see \cite{marshall1979inequalities}). The procedure terminates when the seller and all buyers whose marginal contributions exceed $\sigma$ get pooled at $\sigma$, while every remaining buyer retains her marginal contribution, yielding
\[\pi_s^{LDV}=\sigma\qquad\text{and}\qquad\pi_i^{LDV}=\min\{\sigma,m_i\}.\]
The Sequential Pooling Algorithm and the proof that its output is $\pi^{LDV}$ are provided in Appendix~\ref{app:pooling}.
\end{remark}

The licence example above is special: every player ends up at the common level. The next example shows the general structure, in which some buyers are shaved down to the common level and others keep their marginal contributions.

\begin{example}\label{ex:three-buyer-ldv}
Let $G=\{a,b\}$ and $B=\{1,2,3\}$. Each buyer wants a single good: buyer $1$ values good $a$ at $60$, while buyers $2$ and $3$ value good $b$ at $40$ and $30$.\footnote{Formally, $v_1(A)=60\cdot\1_{\{a\in A\}}$, $v_2(A)=40\cdot\1_{\{b\in A\}}$ and $v_3(A)=30\cdot\1_{\{b\in A\}}$ for every $A\subseteq\{a,b\}$. The induced coalitional values are $W(\{s,1\})=60$, $W(\{s,2\})=40$, $W(\{s,3\})=30$, $W(\{s,1,2\})=100$, $W(\{s,1,3\})=90$, $W(\{s,2,3\})=40$ and $W(N)=100$. Buyers are substitutes: $m_{\{1,2\}}=70\geq70$, $m_{\{1,3\}}=60\geq60$, $m_{\{2,3\}}=40\geq10$ and $m_B=100\geq70$.} The efficient allocation gives $a$ to buyer $1$ and $b$ to buyer $2$, so $W(N)=100$. Removing buyer $1$ loses good $a$'s value, so $m_1=60$. Removing buyer $2$ passes good $b$ to buyer $3$, so $m_2=40-30=10$. Removing buyer $3$ changes nothing, so $m_3=0$. The Vickrey payoff vector is
\[\pi^{VCG}=(30,60,10,0).\]
Buyer $1$ faces no competition for good $a$ and pays nothing. Buyer $2$ pays $30$, buyer $3$'s value. The seller receives $30$, half of what buyer $1$ receives.

The common level solves $\sigma+\min\{\sigma,60\}+\min\{\sigma,10\}+\min\{\sigma,0\}=100$. For $10\leq\sigma\leq60$, this reads $2\sigma+10=100$, so $\sigma=45$, and
\[\pi^{LDV}=(45,45,10,0).\]
Only buyer $1$'s payoff changes: she now pays $15$ for good $a$. Buyer $2$ keeps her marginal contribution $10$, and buyer $3$ keeps $0$. The Lorenz-dominant vector is therefore not an equal split. It equalises only where the caps allow. Characterisation (ii) gives the same level: for $T=\{2,3\}$, the average $(100-10-0)/2$ equals $45$.

The pooling procedure of Remark~\ref{rem:pooling} reaches this vector in one step. The seller ($30$) is poorer than the richest buyer, buyer $1$ ($60$). Pooling the two gives each $(30+60)/2=45$, which is still above the next-richest buyer's payoff of $10$. The procedure therefore stops at $(45,45,10,0)$: a single progressive transfer of $15$ from buyer $1$ to the seller.
\end{example}

Let $n=|B|$ and $\bar x=W(N)/(n+1)$, and let $\pi^{ED}=(\bar x,\bar x,\ldots,\bar x)\in\mathbb{R}^{n+1}$ denote the equal division vector. It is well known that $\pi^{ED}$ Lorenz dominates every payoff vector with the same
total surplus (\cite{hardy1929some}). Every vector in $\mathrm{Core}(N,W)$ has total $W(N)$, so this includes
every core payoff. In other words, $\pi^{ED}$ is already the most egalitarian outcome
imaginable. This has nothing to do with stability, it would be true even if the core was empty.

The concern is that $\pi^{ED}$ need not be stable. An even split can leave some coalition
worse off than it could do on its own, in which case $\pi^{ED}$ simply fails to lie in
$\mathrm{Core}(N,W)$. This leaves one natural question: when does this unconditional
fairness ideal also happen to be stable? The answer is immediate once we ask it this way.
If $\pi^{ED}$ lies in the core at all, it already Lorenz dominates every other point in
that core. By the uniqueness in Theorem \ref{thm:ldv-existence}, it must then be the Lorenz-dominant vector
itself, there is nothing else it could be. So the two questions collapse into one: when
is equal division stable?

\begin{corollary}\label{cor:equal-division}
Suppose buyers are substitutes, and let $\bar x$ and $\pi^{ED}$ be as defined above. Then,

$$\pi^{LDV} = \pi^{ED} \quad \text{if and only if} \quad m_i \ge \bar x \quad \text{for every buyer } i \in B.$$
\end{corollary}

\begin{proof}
($\Leftarrow$) Suppose $m_i \ge \bar x$ for every $i \in B$. Then
\[
\bar x + \sum_{i\in B} \min\{\bar x, m_i\} = \bar x + n\bar x = (n+1)\bar x = W(N),
\]
so $\bar x$ solves the equation defining $\sigma$. By uniqueness of the solution to $g(x) = W(N)$, $\sigma = \bar x$. Theorem
\ref{thm:ldv-existence} then gives $\pi^{LDV}_s = \sigma = \bar x$ and $\pi^{LDV}_i = \min\{\sigma, m_i\} = \bar x$
for every $i \in B$. Hence $\pi^{LDV} = \pi^{ED}$.

\noindent($\Rightarrow$) Suppose $\pi^{LDV} = \pi^{ED}$. Since buyers are substitutes, Theorem \ref{thm:ldv-existence} gives $\pi^{LDV} \in \mathrm{Core}(N,W)$. By the Cap Lemma (Lemma \ref{lem:cap}),
$\pi^{LDV}_i \le m_i$ for every buyer $i$, i.e. $\bar x \le m_i$ for every $i \in B$.
\end{proof}

It is also natural to ask when the Vickrey payoff itself is Lorenz dominant. This occurs precisely when the seller already receives at least as much as every buyer, so no redistribution is needed to reconcile the VCG payoff with fairness. The following corollary characterises this coincidence.

\begin{corollary}\label{cor:vcg-ldv}
Suppose buyers are substitutes. Then,

$$\pi^{LDV} = \pi^{VCG} \quad \text{if and only if} \quad \max_{i \in B} m_i \;\le\; W(N) - \sum_{i \in B} m_i = \pi^{VCG}_s$$
\end{corollary}

\begin{proof}
($\Leftarrow$) Suppose $\max_i m_i \le \pi^{VCG}_s = x^*$. Then $m_i \le x^*$ for every
buyer $i$, so
\[
g(x^*) = x^* + \sum_{i\in B} \min\{x^*, m_i\} = x^* + \sum_{i\in B} m_i
= \Big(W(N) - \sum_{i\in B} m_i\Big) + \sum_{i\in B} m_i = W(N).
\]
By uniqueness of the solution to $g(x) = W(N)$, $\sigma = x^*$. By Theorem \ref{thm:ldv-existence},
$\pi^{LDV}_s = \sigma = x^* = \pi^{VCG}_s$, and for every buyer $i$,
$\pi^{LDV}_i = \min\{\sigma, m_i\} = \min\{x^*, m_i\} = m_i = \pi^{VCG}_i$. Hence $\pi^{LDV} = \pi^{VCG}$.

\noindent($\Rightarrow$) Suppose $\pi^{LDV} = \pi^{VCG}$. Then for every buyer $i$,
$m_i = \pi^{VCG}_i = \pi^{LDV}_i = \min\{\sigma, m_i\}$, which forces $m_i \le \sigma$.
Since $\pi^{LDV}_s = \sigma = \pi^{VCG}_s$, taking the maximum over $i \in B$ gives
$\max_i m_i \le \pi^{VCG}_s$.
\end{proof}

\subsection{Is the substitutes condition necessary?}\label{subsec:converse}
Theorem~\ref{thm:ldv-existence} shows that the buyers-are-substitutes condition is sufficient for the existence of a Lorenz-dominant payoff in the core. It is natural to ask whether it is also necessary. The answer is no. If the condition fails, the core may or may not contain a Lorenz-dominant payoff. We show both possibilities below. The detailed reasoning and proofs are provided in Appendix~\ref{app:examples}.

Recall the auction in Example~\ref{ex:vcg-core-failure}, with three buyers and two goods. The maximum surplus generated by each group of buyers is
\[
\begin{array}{c|cccccccc}
S&\varnothing&\{1\}&\{2\}&\{3\}&\{1,2\}&\{1,3\}&\{2,3\}&B\\\hline
U(S)&0&2&2&2&2&2&4&4
\end{array}
\]
Buyers are not substitutes in this example: $m_{\{2,3\}}=2<4=m_2+m_3$. Nevertheless, $x^*=(2,0,1,1)$ is Lorenz dominant in the core, with coordinates ordered as $(s,1,2,3)$. Thus, the buyers-are-substitutes condition is not necessary for the existence of a Lorenz-dominant vector in the core.

For completeness, the next example shows that, when buyers are not substitutes, a Lorenz-dominant payoff need not exist either.

\begin{example}
\label{ex:no-substitutes-no-ldv}
Let $B=\{1,2,3\}$.  Suppose that the maximum surplus generated by each
group of buyers is
\[
\begin{array}{c|cccccccc}
S
    & \varnothing & \{1\} & \{2\} & \{3\}
    & \{1,2\} & \{1,3\} & \{2,3\} & B \\
\hline
U(S)
    & 0 & 7 & 1 & 9 & 7 & 11 & 9 & 12.
\end{array}
\]
These values can be generated by a combinatorial auction with three indivisible
goods.\footnote{Let the goods be $a,b,c$.  One set of normalized, monotone
valuations generating the coalitional values in Example \ref{ex:no-substitutes-no-ldv} is $v_1(A)=\max\bigl\{4\mathbf 1_{\{a\in A\}},
                   7\mathbf 1_{\{b,c\}\subseteq A}\bigr\},
\;
v_2(A)=\mathbf 1_{\{c\in A\}},
\;
v_3(A)=\max\bigl\{7\mathbf 1_{\{b\in A\}},
                   9\mathbf 1_{\{a,c\}\subseteq A}\bigr\}$,
for every $A\subseteq\{a,b,c\}$.} Buyers are not substitutes: $m_{\{1,2\}}=3<4=m_1+m_2$. The core contains the two Lorenz-maximal payoffs
\[x^A=\left(\frac92,\frac52,\frac12,\frac92\right)\qquad\text{and}\qquad x^B=(5,2,1,4),\]
with coordinates again ordered as $(s,1,2,3)$. These payoffs are Lorenz incomparable, and the core contains no Lorenz-dominant payoff.
\end{example}
Therefore, without the buyers-are-substitutes condition, either outcome is possible: the core may have a Lorenz-dominant payoff, as in Example~\ref{ex:vcg-core-failure}, or it may have several Lorenz-maximal payoffs and no Lorenz-dominant payoff, as in Example~\ref{ex:no-substitutes-no-ldv}.

\subsection{Conditions on the buyer valuations that guarantee existence of LDV}
The existence result assumes that buyers are substitutes. But $W$ is not the most primitive object: it is derived from each buyer's own valuation $v_i$ over bundles of goods. It is worth asking what these valuations need to look like for the induced coalitional value function to satisfy the buyers-are-substitutes condition.

The standard answer in the literature is the gross-substitutes condition, introduced by \citet{kelso1982job} for a labor market in which firms hire sets of workers, and later adapted to indivisible goods by \citet{gul1999walrasian}. Informally, a buyer's valuation satisfies gross substitutes if raising the price of some goods never reduces her demand for a good whose price has not changed. \cite{ausubel2002ascending,ausubel2006lovely} imply that if every buyer's valuation satisfies gross substitutes, then buyers are substitutes in the induced auction game.
\begin{corollary}
If every buyer's valuation $v_i$ satisfies the gross-substitutes condition, then $\Core(N,W)$ contains a Lorenz-dominant vector $\pi^{LDV}$.
\end{corollary}
\begin{proof}
If every buyer's valuation satisfies gross substitutes, \citet{ausubel2002ascending} imply that buyers are substitutes. Hence, by Theorem~\ref{thm:ldv-existence}, a Lorenz-dominant vector exists in $\Core(N,W)$.
\end{proof}

\section{A Non-Cooperative Implementation of the Lorenz-Dominant Vector}
\label{sec:implementation}
This section first explains why an efficient, DSIC auction cannot implement $\pi^{LDV}$ in general. It then presents a complete-information bargaining game, the seller-proposer game, whose every subgame-perfect equilibrium yields $\pi^{LDV}$ (Theorem~\ref{thm:seller-proposer-implementation}).
\subsection{Core selection and the limits of truthful implementation}
Can an auction implement $\pi^{LDV}$ when valuations are privately known? \citet{day2008core} study auctions that maximise reported surplus and select payoffs from the reported core. This does not ensure truthful reporting.

Suppose the domain of valuations is regular, that is, a product of convex sets, one for each buyer, each containing the zero valuation. By \citet{holmstrom1979groves}, efficiency and dominant-strategy incentive compatibility (DSIC) require Groves payments: $p_i=h_i(v_{-i})-\sum_{j\ne i}v_j(X_j^*)$, where $h_i$ is independent of buyer $i$'s report.

Stability requires the true payoff vector to lie in the core at every profile. The singleton constraints give $x_j\geq0$ for every player, ensuring ex post individual rationality. Moreover, every buyer's core payoff is bounded above by her marginal contribution, $x_i\leq m_i$ \citep{ausubel2002ascending}. The other buyers can retain their bundles when $i$ is absent, so $W(N\setminus\{i\})\geq\sum_{j\ne i}v_j(X_j^*)$. Thus $m_i\leq v_i(X_i^*)$, and $p_i=v_i(X_i^*)-x_i\geq0$. Neither individual rationality nor nonnegative payments is an additional assumption.

Fix $v_{-i}$ and set $v_i=0$. Individual rationality and nonnegative payments force $p_i=0$. Efficiency then gives $h_i(v_{-i})=W(N\setminus\{i\})$. Since $h_i$ is independent of $v_i$, these are the VCG payments at every profile. Conversely, VCG is efficient and DSIC, with $x_i=m_i\geq0$ and $p_i\geq0$ by the same feasibility argument. If buyers are substitutes at every profile, VCG is also stable. Under this condition, efficiency, DSIC and stability therefore characterise VCG.
\begin{corollary}\label{cor:private-ldv}
Suppose the domain of valuations is regular and buyers are substitutes in $W_v$ at every valuation profile $v$. There exists an efficient, DSIC auction implementing $\pi^{LDV}(v)$ at every profile if and only if
\[
\max_{i\in B}m_i(v)\leq W_v(N)-\sum_{i\in B}m_i(v)=\pi_s^{VCG}(v)\qquad\text{for every }v,
\]
where $m_i(v)=W_v(N)-W_v(N\setminus\{i\})$.
\end{corollary}
\begin{proof}
Necessity follows from the preceding characterisation and Corollary~\ref{cor:vcg-ldv}. Conversely, the inequality implies $\pi^{LDV}(v)=\pi^{VCG}(v)$ at every profile, so VCG implements $\pi^{LDV}$.
\end{proof}
The buyers-are-substitutes condition does not ensure this coincidence. We therefore take $W$ as common knowledge and, in the spirit of the Nash program, construct a non-cooperative bargaining game implementing its Lorenz-dominant vector.

\subsection{The Seller-Proposer Game}\label{subsec:seller-proposer-game}
Under complete information, implementation by itself is easy. A game in which
the seller receives zero unless she proposes $\pi^{LDV}$, and a rejecting buyer
receives zero, implements $\pi^{LDV}$ in subgame-perfect equilibrium. Such a game,
however, writes the target into its rules: the designer must compute
$\pi^{LDV}$ in advance and could then simply impose it. It also enforces the
target by destroying surplus off the equilibrium path, a threat the players
would renegotiate away. In the spirit of the Nash program, we therefore require
the rules of the game to (i) make no reference to the target vector, (ii) allow
the seller to propose any feasible division, and (iii) distribute the entire
surplus after every history. The game below satisfies all three. Its rules use
only the buyers' marginal contributions, the remaining surplus and the seller's
own proposal, and the Lorenz-dominant vector emerges only as the equilibrium
outcome.

Our game is in the spirit of the recursive bargaining constructions of \citet{perez2001bidding} and \citet{ju2009implementing}. We use proposals, sequential responses and continuation after rejection, with the seller as the sole proposer and an exit payment designed to implement the Lorenz-dominant vector.

Throughout, the allocation of the goods is taken as given. Before the game begins, an efficient allocation $X^*$ is fixed, and buyer $i$ receives the bundle $X_i^*$ whatever happens in the game. Bargaining determines only how the surplus $W(N)=\sum_{i\in B}v_i(X_i^*)$ is divided, that is, the payments that buyers make to the seller. We return to this interpretation after describing the game.

A node is a pair $(C,w)$, where $C\subseteq B$ is the set of active buyers and $w\geq0$ is the remaining surplus. The seller remains active at every node $(C,w)$, and the initial node is $(B,W(N))$. Each $m_i$ remains buyer $i$'s marginal contribution to the original grand coalition; it is not recomputed after an exit (see the discussion following the description of the game). Fix an order of the buyers before the game begins. At every node, the active buyers respond to the proposal of seller according to its restriction to $C$.

At each node $(C,w)$, we seek to implement the division obtained by applying the formula in Theorem~\ref{thm:ldv-existence} to the active buyers and the remaining surplus. Let $\sigma_{C,w}$ be the unique nonnegative solution of
\begin{equation}\label{eq:node-common-level}
x+\sum_{i\in C}\min\{m_i,x\}=w,
\end{equation}
and define the target payoffs by
\begin{equation}\label{eq:node-target-payoffs}
\ell_s(C,w)=\sigma_{C,w},\qquad\ell_i(C,w)=\min\{m_i,\sigma_{C,w}\}\quad\text{for every }i\in C.
\end{equation}
The common level is well defined because the left-hand side of \eqref{eq:node-common-level} is continuous and strictly increasing from zero to infinity. The target payoffs sum to $w$. At the initial node, $\sigma_{B,W(N)}=\sigma$, so these payoffs coincide with $\pi^{LDV}$. Following arbitrary earlier exits, however, the targets at a continuation node need not equal the corresponding components of the original Lorenz-dominant vector.

The target division is preserved when a buyer leaves with her target payoff: subtracting $\ell_i(C,w)$ from the remaining surplus leaves the same common level for the other players. Conversely, a buyer who leaves with less than her target raises the common level of those who remain, and in particular the seller's, because the common level increases with the remaining surplus (Lemma~\ref{lem:node-common-level}). The exit rule must therefore protect a buyer against being induced to leave with less than her target, while allowing the seller to propose a division under which rejection preserves her own target payoff. The rule may use only the seller's proposal, the marginal contributions and the remaining surplus. We construct it from two claims, each using the seller's proposed share as a benchmark. Throughout this discussion we use the values of Example~\ref{ex:two-buyer-seller-game} below: $W(N)=100$, $m_1=60$, $m_2=20$ and $\pi^{LDV}=(40,40,20)$.

\paragraph{The capped claim.} Suppose that the game is at a node $(C,w)$ and that the seller proposes to retain $z$. Her proposal reveals the share she claims for herself. A natural benchmark is equal treatment: the rejecting buyer may claim the same amount. This mirrors alternating-offers bargaining between equally patient players, where neither side has more bargaining power and the split tends to equality \citep{rubinstein1982perfect}. The buyer cannot, however, claim more than she adds to the surplus: no core payoff gives buyer $i$ more than $m_i$ (Lemma~\ref{lem:cap}). This gives the capped marginal-contribution claim
\begin{equation}\label{eq:capped-claim}A_i(z)=\min\{m_i,z\}.\end{equation}
If the seller proposes her target share $z=\sigma_{C,w}$, the capped claim equals the buyer's target $\ell_i(C,w)$.

The capped claim alone, however, lets the seller squeeze a buyer. Suppose a rejecting buyer received only $A_i(z)$. The seller could propose $(30,0,70)$, claiming only $z=30$ and offering buyer $1$ nothing. Buyer $1$ would reject and leave with $A_1(30)=\min\{60,30\}=30$, below her target of $40$. The game would continue at $(\{2\},70)$, whose common level is $\sigma_{\{2\},70}=50$. The seller could secure this amount, for instance by proposing $(50,20)$: buyer $2$ is willing to accept, since rejecting would give her only $A_2(50)=20$. By understating her claim now, the seller makes the exit cheap and recovers the surplus later, ending with $50>40$.

\paragraph{The residual claim.} The buyer's reply to such a proposal is to hold the seller to her own claim. If the seller claims only $z$, then by her own standard every other buyer $h$ is owed at most $\min\{m_h,z\}$, and the rest belongs to buyer $i$. This gives the residual claim
\begin{equation}\label{eq:residual-claim}
R_i(C,z;w)=w-z-\sum_{h\in C\setminus\{i\}}\min\{m_h,z\}.
\end{equation}
In the squeeze above, $R_1(\{1,2\},30;100)=100-30-20=50$. Under this claim, buyer $1$ leaves with $50$, and the continuation node $(\{2\},50)$ has common level $30$. The squeeze now backfires: the seller ends with less than her target.

The residual claim alone is not enough either, because it can be lowered by overstating $z$. If the seller claims $z=60$, then $R_1(\{1,2\},60;100)=100-60-20=20$. A buyer paid only her residual claim would leave with $20$, again below her target, and the continuation node $(\{2\},80)$ has common level $60>40$. At $z=60$, however, the capped claim is $A_1(60)=60$.

Each claim therefore guards one side. Understating $z$ lowers the capped claim but raises the residual claim; overstating $z$ lowers the residual claim but raises the capped claim. A rejecting buyer receives the larger of the two. The first claim applies the proposed common level directly to buyer $i$; the second applies it to everyone else and asks what remains for her. Neither claim depends on the amount offered to buyer $i$.

\begin{remark}[Monotonicity and intersection of the two claims]\label{rem:claim-monotonicity}
Fix a node $(C,w)$ and a buyer $i\in C$. The capped claim $A_i(z)$ is weakly increasing in $z$, whereas the residual claim $R_i(C,z;w)$ is strictly decreasing: a larger $z$ reserves more for the seller and weakly more for every other buyer. Moreover,
\[A_i(z)=R_i(C,z;w)\quad\Longleftrightarrow\quad z+\sum_{h\in C}\min\{m_h,z\}=w.\]
Thus, the two claims intersect uniquely at the target common level, where
\[A_i(\sigma_{C,w})=R_i(C,\sigma_{C,w};w)=\ell_i(C,w).\]
For $z<\sigma_{C,w}$, the residual claim exceeds the buyer's target payoff; for $z>\sigma_{C,w}$, the capped claim is at least that payoff.
\end{remark}

\begin{figure}[htbp]
\centering
\begin{tikzpicture}[x=1.05cm,y=.55cm,>=stealth,font=\small]
\begin{scope}
\draw[->] (0,-.7)--(0,5.7) node[above]{Claim};
\draw[->] (0,0)--(4.2,0) node[right]{$z$};
\draw[thick] (0,0)--(2.6,2.6)--(4,2.6) node[above left]{$A_i(z)$};
\draw[thick] (0,5)--(1.5,2)--(4,-.5);
\node[right] at (.65,4.2){$R_i(C,z;w)$};
\draw[dashed] (0,1.75)--(1.75,1.75)--(1.75,0);
\node[left] at (0,5){$w$};\node[below left] at (0,0){$0$};
\node[left] at (0,1.75){$\ell_i(C,w)$};\node[below] at (1.75,0){$\sigma_{C,w}$};
\node at (2,-1.65){(a) $m_i>\sigma_{C,w}$};
\end{scope}
\begin{scope}[xshift=7.3cm]
\draw[->] (0,-.7)--(0,5.7) node[above]{Claim};
\draw[->] (0,0)--(4.2,0) node[right]{$z$};
\draw[thick] (0,0)--(1.2,1.2)--(4,1.2) node[above left]{$A_i(z)$};
\draw[thick] (0,5)--(1.8,1.4)--(4,-.8);
\node[right] at (.65,4.2){$R_i(C,z;w)$};
\draw[dashed] (0,1.2)--(2,1.2)--(2,0);
\node[left] at (0,5){$w$};\node[below left] at (0,0){$0$};
\node[left] at (0,1.2){$\ell_i(C,w)$};\node[below] at (2,0){$\sigma_{C,w}$};
\node at (2,-1.65){(b) $m_i<\sigma_{C,w}$};
\end{scope}
\end{tikzpicture}
\caption{The two claims at a node $(C,w)$. In both panels, they intersect at $(\sigma_{C,w},\ell_i(C,w))$. The intersection lies on the rising part of the capped claim in panel (a) and on its flat part in panel (b); when $m_i=\sigma_{C,w}$, it lies at the kink. Kinks in the residual claim occur when other buyers' caps bind. The curves are schematic.}\label{fig:claims}
\end{figure}
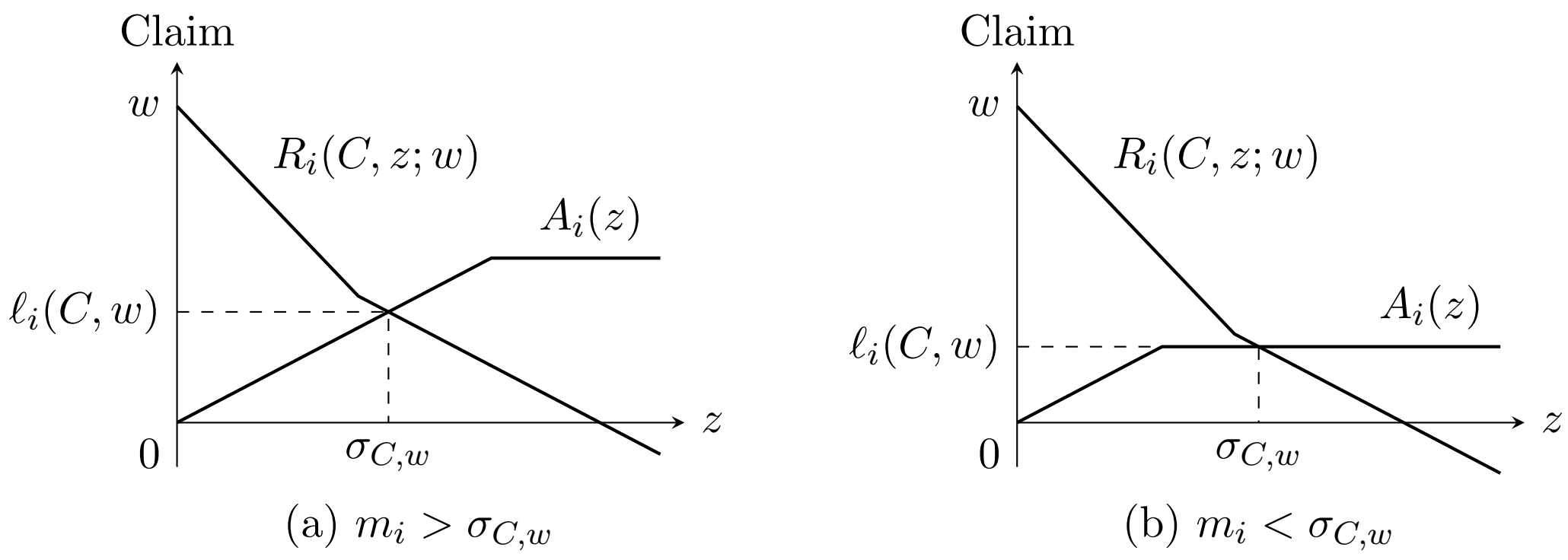

Figure~\ref{fig:claims} illustrates why the larger claim provides the required protection. Whatever share the seller proposes for herself, a buyer who can leave with the larger claim receives at least her target payoff. Yet the seller can make both claims equal to that target by proposing $z=\sigma_{C,w}$. Under the target proposal, acceptance gives each player her target, while rejection pays the departing buyer exactly her target and preserves those of the remaining players. The game below incorporates this payment into its response stage.

\noindent\textbf{Seller-proposer game}

\noindent The seller-proposer game, denoted by $\Gamma^S$, begins at $(B,W(N))$ and proceeds recursively as follows. At any node $(C,w)$ with $C\ne\varnothing$, the game has two stages.
\begin{enumerate}
\item \emph{Proposal.} The seller proposes a feasible division of the remaining surplus,
\[\left(z,(x_i)_{i\in C}\right),\]
satisfying
\[z+\sum_{i\in C}x_i=w,\qquad z\geq0,\qquad x_i\geq0\quad\text{for every }i\in C.\]
Here, $z$ is the amount proposed for the seller and $x_i$ is the amount proposed for buyer $i$.
\item \emph{Responses.} After observing the entire proposal, the buyers in $C$ respond sequentially according to the fixed order. Each buyer observes the responses made before her and chooses either to accept or to reject.

If every buyer accepts, the proposal is implemented. The seller receives $z$, every active buyer $i\in C$ receives $x_i$, and the game ends.

If buyer $j$ is the first buyer to reject, the response stage ends immediately and the proposal is discarded. None of the proposed amounts is paid. Buyer $j$ instead receives the larger of her two claims,
\begin{equation}\label{eq:exit-payment}
e_j(C,z;w)=\max\{A_j(z),R_j(C,z;w)\},
\end{equation}
from the remaining surplus and exits the game permanently. This payment depends on the seller's proposed share and the two claims, not on the amount offered to buyer $j$.

The other buyers remain active, including any buyer who accepted before buyer $j$ rejected. The game then restarts with a new proposal by the seller at the node
\[\left(C\setminus\{j\},w-e_j(C,z;w)\right).\]
\end{enumerate}
Payments received by buyers who have already exited are final. Such buyers take no further actions and do not share in any subsequent division of the remaining surplus. If the game reaches a node $(\varnothing,w)$, the seller receives the entire remaining surplus $w$ and the game ends. Since every rejection removes one active buyer, the game terminates after at most $|B|$ rejections. There is no discounting between successive nodes.

\subsection{The Main Implementation Result}\label{subsec:implementation-result}
Our second main result shows that the seller-proposer game delivers exactly the Lorenz-dominant division. Throughout the implementation analysis, we consider pure strategies.
\begin{theorem}\label{thm:seller-proposer-implementation}
Suppose buyers are substitutes. Then the seller-proposer game $\Gamma^S$ fully implements the Lorenz-dominant vector in subgame-perfect equilibrium: a subgame-perfect equilibrium exists, and the payoff vector in every subgame-perfect equilibrium is $\pi^{LDV}$.
\end{theorem}
The seller makes every proposal, yet she cannot obtain more than her Lorenz-dominant payoff $\sigma$: any buyer can reject and leave with at least her own Lorenz-dominant payoff. The remainder of this subsection explains how the game should be read as an auction and illustrates it with an example. Subsection~\ref{subsec:equilibrium-proof-idea} sketches the proof, and Appendix~\ref{app:implementation} gives the details.

\paragraph{Interpretation: settling payments for a fixed efficient allocation.} The game does not reallocate goods. Every buyer receives her bundle $X_i^*$ in the efficient allocation, and every amount in the game is a net payoff. A buyer who ends the game with payoff $x_i$ pays the seller
\begin{equation}\label{eq:settlement-payment}
p_i=v_i(X_i^*)-x_i.
\end{equation}
In particular, a rejecting buyer $j$ does not walk away from the auction. She receives her bundle $X_j^*$ and settles her payment at $p_j=v_j(X_j^*)-e_j(C,z;w)$, which is final. We call $e_j$ an exit payment because it is paid to buyer $j$ out of the surplus; in auction terms it is her net utility, not a transfer from the seller.

The remaining surplus has a simple accounting meaning. At every node $(C,w)$ reached in the game,
\begin{equation}\label{eq:remaining-surplus-accounting}
w=\sum_{i\in C}v_i(X_i^*)+\sum_{k\in B\setminus C}p_k.
\end{equation}
The first term is the value of the bundles of the active buyers, whose payments are still to be settled. The second term is the total payment already committed by the buyers who have exited. Equation~\eqref{eq:remaining-surplus-accounting} holds at the initial node, where $w=W(N)$ and no buyer has exited. If buyer $j$ exits from $(C,w)$, then $w-e_j=\sum_{i\in C\setminus\{j\}}v_i(X_i^*)+\sum_{k\in B\setminus C}p_k+p_j$, so it continues to hold at the next node. Consequently, whenever the game ends, the seller's payoff equals the total payment she collects from the buyers. If all buyers accept a proposal at $(C,w)$, the seller receives $z=w-\sum_{i\in C}x_i=\sum_{i\in B}p_i$. At a node $(\varnothing,w)$, she receives $w=\sum_{i\in B}p_i$.

The outcome of the game is therefore an auction outcome: the efficient allocation $X^*$ together with a payment vector $p$. In equilibrium these payments are nonnegative. By Theorem~\ref{thm:seller-proposer-implementation}, every buyer ends with $\pi_i^{LDV}\leq m_i$, and $m_i\leq v_i(X_i^*)$ because the other buyers can keep their bundles when buyer $i$ is absent (Section~\ref{sec:implementation}). Hence $p_i\geq0$ for every buyer, and every buyer's payoff is nonnegative. Off the equilibrium path, the rules allow an exit payment above $v_j(X_j^*)$, which would mean a transfer from the seller to buyer $j$; such payments never occur in equilibrium. If several allocations are efficient, any one can be fixed: $W(N)$ and the marginal contributions do not depend on the choice, so neither do the equilibrium payoffs.

This interpretation also explains why the marginal contributions are not recomputed after an exit. A continuation node is not a smaller auction among the active buyers. Buyers who have exited still hold their bundles, so the coalitional values of the remaining buyers, such as $W(N\setminus\{j\})$, do not describe the surplus that remains to be divided. Each $m_i$ is buyer $i$'s VCG payoff in the original auction and, by the Cap Lemma, the largest payoff she can receive in any core outcome. It is fixed at the outset as a cap on what she can claim, and the game keeps it unchanged at every node.

The following example illustrates the game with one seller and two buyers.
\begin{example}
\label{ex:two-buyer-seller-game}
Consider a seller and two buyers. Suppose that the relevant coalitional
values are
\[
    W(N)=100,
    \qquad
    W(N\setminus\{1\})=40,
    \qquad
    W(N\setminus\{2\})=80.
\]
These coalitional values can, for example, arise from a simple
two-good combinatorial auction.\footnote{Let the goods be $a$ and $b$.
Buyer $1$ has the additive valuation
$v_1(\{a\})=60$, $v_1(\{b\})=20$, and
$v_1(\{a,b\})=80$, while buyer $2$ has
$v_2(\{a\})=0$, $v_2(\{b\})=40$, and
$v_2(\{a,b\})=40$. The efficient allocation gives good $a$ to buyer
$1$ and good $b$ to buyer $2$, generating a total surplus of $100$.}
Buyers are substitutes: with two buyers, the only nontrivial inequality is
\[
m_{\{1,2\}}=W(N)-W(\{s\})=100\geq80=60+20=m_1+m_2.
\]
Theorem~\ref{thm:ldv-existence} therefore applies. The buyers' marginal contributions are therefore
\[
    m_1
    =
    W(N)-W(N\setminus\{1\})
    =
    100-40
    =
    60
\]
and
\[
    m_2
    =
    W(N)-W(N\setminus\{2\})
    =
    100-80
    =
    20.
\]
At the initial node $(\{1,2\},100)$, the common target level solves
\[
    \sigma+\min\{60,\sigma\}+\min\{20,\sigma\}=100.
\]
The solution is $\sigma=40.$ Thus, the target payoff vector coincides with the Lorenz-dominant vector, $\pi^{LDV}=(40,40,20).$
Suppose that the buyers respond in the order $1,2$, and the seller
proposes the target vector $(40,40,20).$
When the seller's proposed share is $z=40$, buyer $1$'s exit payment is
\begin{align*}
    e_1(\{1,2\},40;100)
    &=
    \max\left\{
        \min\{60,40\},
        100-40-\min\{20,40\}
    \right\}
    =40.
\end{align*}
Similarly, buyer $2$'s exit payment is
\begin{align*}
    e_2(\{1,2\},40;100)
    &=
    \max\left\{
        \min\{20,40\},
        100-40-\min\{60,40\}
    \right\}
    =20.
\end{align*}
Each buyer is therefore indifferent between accepting and rejecting.
If both buyers accept, the game ends with the payoff vector $(40,40,20).$
The seller need not, however, propose this particular division among
the buyers. For example, consider the proposal $(40,10,50).$
The seller still asks for her target share, but she offers only $10$ to
buyer $1$ and $50$ to buyer $2$. If buyer $2$ is reached, she accepts
because $50$ is greater than her exit payment of $20$. Buyer $1$
therefore knows that accepting gives her $10$, whereas rejecting gives
her $40$. She rejects and exits with $40$.
The game then continues at the reduced node $(\{2\},60).$
At this node, the common level solves
\[
    \sigma_{\{2\},60}
    +
    \min\{20,\sigma_{\{2\},60}\}
    =
    60,
\]
and hence $\sigma_{\{2\},60}=40.$ The seller can therefore propose $(40,20)$
to buyer $2$. Whether buyer $2$ accepts this proposal or rejects and
exits with $20$, the final payoff vector in the original game is again $(40,40,20).$

In terms of the auction, good $a$ goes to buyer $1$ and good $b$ to buyer $2$ throughout. When buyer $1$ rejects and exits with $40$, she receives good $a$ and pays $60-40=20$. The surplus of $60$ at the node $(\{2\},60)$ is buyer $2$'s value $40$ for good $b$ plus buyer $1$'s payment of $20$, as in \eqref{eq:remaining-surplus-accounting}. It is not $W(N\setminus\{1\})=40$, because buyer $1$ has not left the auction. Buyer $2$ ends with $20$ and pays $40-20=20$. The seller's revenue is $20+20=40$, her Lorenz-dominant payoff.

This example illustrates the two possible ways in which the target
vector can be reached. The seller may make the target proposal and have
it accepted immediately, or she may choose a different division among
the buyers, cause a rejection, and reach the same payoffs through the
continuation game. Thus, different optimal proposals by the seller can yield the same Lorenz-dominant payoff vector.
\end{example}

\subsection{Equilibrium and proof idea}\label{subsec:equilibrium-proof-idea}
We first show that the seller can propose the target division and have it accepted by all buyers in equilibrium. We then explain why every equilibrium gives the same payoffs. At a node $(C,w)$, the targets are $\sigma_{C,w}$ for the seller and $\ell_i(C,w)$ for buyer $i$. At the initial node, these are the Lorenz-dominant payoffs.

\paragraph{An equilibrium with immediate agreement.} At every node $(C,w)$, let the seller propose the target division,
\begin{equation}\label{eq:target-proposal}
P^{\mathrm{tar}}(C,w)=\left(\sigma_{C,w},(\ell_i(C,w))_{i\in C}\right),
\end{equation}
and let every buyer accept it. We must also specify how buyers respond to other proposals. Start with the last buyer and work backwards. Each buyer compares what she gets by rejecting with what she gets by accepting. If all later buyers accept, acceptance gives her the proposed amount. If a later buyer rejects, she remains in the game and receives her target at the reduced node. Choose the action giving her the larger payoff, and choose acceptance when the two payoffs are equal. Define these strategies first at nodes with no buyers, then with one buyer, and so on, so that play after every rejection is already specified.

We verify by induction that these strategies form a subgame-perfect equilibrium. There is nothing to prove when no buyers remain. Suppose the strategies form an equilibrium at all nodes with fewer buyers than $C$. At the target proposal, buyer $i$ gets $\ell_i(C,w)$ whether she accepts or rejects, so acceptance is optimal. At every other proposal, buyers' responses are optimal by the construction above.

The seller also has no profitable deviation. If all buyers accept an alternative proposal, each must be offered at least her exit payment and hence at least her target. This leaves at most $\sigma_{C,w}$ for the seller. If buyer $j$ rejects instead, she receives at least $\ell_j(C,w)$. Paying exactly this amount preserves the seller's target at the reduced node; paying more lowers it. By the induction hypothesis, the seller receives her target at that node, so rejection cannot give her more than $\sigma_{C,w}$ either. The target proposal is therefore optimal. This completes the induction and establishes an equilibrium in which the initial proposal $\pi^{LDV}$ is accepted by everyone.

\paragraph{Why other equilibria give the same payoffs.} Suppose, inductively, that every equilibrium at a node with fewer buyers gives the target payoffs. The seller can secure $\sigma_{C,w}$ by proposing the target division: acceptance gives her this amount directly, and any rejection pays the departing buyer exactly her target, leaving the seller's target unchanged. No other proposal can give the seller more, by considering acceptance and rejection as above. Thus, her payoff in every equilibrium is exactly $\sigma_{C,w}$ (Lemmas~\ref{lem:seller-target-guarantee} and~\ref{lem:seller-upper-bound}).

If all buyers accept, each must receive at least her target. Since the seller receives her target and the targets already sum to $w$, every buyer must receive exactly her target. If a buyer rejects, her exit payment must also equal her target. It cannot be smaller, by the exit rule, and it cannot be larger, because this would lower the seller's continuation payoff below $\sigma_{C,w}$ (Lemmas~\ref{lem:responder-floor} and~\ref{lem:floored-exit-continuation}). The remaining players' targets are then unchanged, and the induction hypothesis gives them these payoffs in the continuation game. Starting from the initial node, every buyer who exits therefore receives her Lorenz-dominant payoff, and the final payoff vector is $\pi^{LDV}$. The supporting lemmas and the full induction argument are in Appendix~\ref{app:implementation}.

\begin{remark}[Multiple equilibria, the same payoffs]\label{rem:multiple-equilibria}
Theorem~\ref{thm:seller-proposer-implementation} gives a unique equilibrium payoff vector, but the game can have multiple subgame-perfect equilibria. The equilibrium with immediate agreement described above is one of them. A larger class arises when, at a node $(C,w)$, the seller proposes her target share for herself and any division of the remaining surplus among the buyers. Such a proposal has the form
\[\left(\sigma_{C,w},(x_i)_{i\in C}\right),\quad x_i\geq0,\quad\sum_{i\in C}x_i=w-\sigma_{C,w}.\]
If the proposed division differs from the target vector, some buyer must reject. Every rejecting buyer receives exactly her target, and her exit leaves the targets of the remaining players unchanged. With equilibrium play thereafter, the seller therefore receives $\sigma_{C,w}$ and every buyer receives her target, whatever division was initially proposed. Each such proposal can be supported in a subgame-perfect equilibrium. At the initial node, this gives equilibria in which the seller proposes her Lorenz-dominant share but divides the buyers' surplus in different ways, as illustrated by $(40,40,20)$ and $(40,10,50)$ in Example~\ref{ex:two-buyer-seller-game}.

These are not the only possible equilibria. The seller may also propose more than her Lorenz-dominant share and still receive that payoff after her proposal is rejected. The following two cases use the values in Example~\ref{ex:two-buyer-seller-game}.

First, return to the continuation node $(\{2\},60)$ in Example~\ref{ex:two-buyer-seller-game}. The target division is $(40,20)$. If the seller instead proposes $(50,10)$, buyer 2's exit payment is
\[e_2(\{2\},50;60)=\max\{20,60-50\}=20.\]
Buyer 2 rejects the offer of 10 and leaves with 20. The seller receives the remaining 40, although she proposed 50 for herself.

The same can happen with both buyers present. Keep the coalitional values of Example~\ref{ex:two-buyer-seller-game}, so that $m_1=60$, $m_2=20$ and $\pi^{LDV}=(40,40,20)$, but fix the response order as $2,1$ before the game begins. Consider the proposal $(z,x_1,x_2)=(50,50,0)$. The exit payments are
\[
e_2(\{1,2\},50;100)=\max\{20,0\}=20,\quad
e_1(\{1,2\},50;100)=\max\{50,30\}=50.
\]
If buyer 1 is reached, accepting and rejecting both give her 50; let her accept. Buyer 2 then strictly prefers rejecting and receiving 20 to accepting and receiving 0. Her exit leaves the node $(\{1\},80)$, where the target division is $(40,40)$. The seller proposes this division and buyer 1 accepts. The final payoffs are again $(40,40,20)$.

In both cases the seller receives 40, which she cannot improve upon in equilibrium. These proposals can therefore be supported in subgame-perfect equilibrium, with responses to all other proposals specified as above. The buyer who exits receives exactly her target, preserving the seller's payoff. In the two-buyer case, buyer 1's initial exit payment is above her target, but she never takes it: buyer 2 rejects first and the proposal is discarded.
\end{remark}

\appendix
\section{Proof of Theorem \ref{thm:ldv-existence}}\label{app:existence}
This appendix gives two proofs of Theorem~\ref{thm:ldv-existence}. Appendix~\ref{app:existence-direct} gives a self-contained direct proof. Appendix~\ref{app:existence-convex} shows that, when buyers are substitutes, the core coincides with the core of a convex game, and derives the theorem from \citet{dutta1989concept}.

\subsection{A direct proof}\label{app:existence-direct}
First note that $\sigma$ is well defined. The function $g(x)=x+\sum_{i\in B}\min\{x,m_i\}$ is strictly increasing in $x$ as the term $x$ is strictly increasing, and each term $\min\{x,m_i\}$ is non-decreasing, so their sum is strictly increasing. Since $g(0)=0\leq W(N)$ and $g(x)\to\infty$ as $x\to\infty$, continuity of $g$ and the intermediate value theorem give a unique solution $\sigma$ to $g(x)=W(N)$.

We now establish two lemmas before proving the theorem.
\begin{lemma}[Cap Lemma]\label{lem:cap}
For every $x\in\Core(N,W)$ and every buyer $i\in B$,
\[x_i\leq m_i.\]
\end{lemma}
Cap Lemma tells us that the payoff of any buyer in the core is capped by her marginal contribution $m_i$ to the combinatorial auction. For proof see \citep{ausubel2002ascending}.
\begin{lemma}[Transfer Lemma]\label{lem:transfer}
Let $x\in\Core(N,W)$, let $i\in B$, and let $t\in[0,x_i]$. Define $x'$ by
\[x'_i=x_i-t,\qquad x'_s=x_s+t,\qquad x'_j=x_j\quad\forall j\ne i\]
Then $x'\in\Core(N,W)$.
\end{lemma}
\begin{proof}
Since $\sum_{j\in N}x'_j=\sum_{j\in N}x_j=W(N)$, it suffices to check the coalitional constraints. Let $C\subseteq N$. If $s\notin C$, then $W(C)=0$, and since $x'_j\geq0$ for all $j$ ($t\leq x_i$ ensures $x'_i\geq0$), we have $\sum_{j\in C}x'_j\geq0=W(C)$. If $s\in C$ and $i\in C$, then $x'_s+x'_i=x_s+x_i$, so $\sum_{j\in C}x'_j=\sum_{j\in C}x_j\geq W(C)$. If $s\in C$ and $i\notin C$, then $x'_s=x_s+t\geq x_s$, so $\sum_{j\in C}x'_j\geq\sum_{j\in C}x_j\geq W(C)$. In every case the constraint for $C$ holds, so $x'\in\Core(N,W)$.
\end{proof}
Transfer lemma tells us that when $x\in\Core(N,W)$, transferring some amount $t\leq x_i$ from buyer $i$ to the seller doesn't move the modified payoff vector $x'$ out of the core. We now begin with the proof of the theorem.

\begin{proof}[\textbf{Proof of Theorem \ref{thm:ldv-existence}}]
Let $\sigma$ be the unique nonnegative solution to
\begin{equation}\label{eq:common-level-proof}
x+\sum_{i\in B}\min\{x,m_i\}=W(N)
\end{equation}
whose existence and uniqueness were established above. Define
\[\pi_s^{LDV}=\sigma,\qquad\pi_i^{LDV}=\min\{\sigma,m_i\}\quad\text{for every }i\in B.\]
Partition the buyers into
\[I=\{i\in B:m_i<\sigma\},\qquad S=\{i\in B:m_i\geq\sigma\}.\]
Substituting $x=\sigma$ in \eqref{eq:common-level-proof} we get
\begin{equation}\label{eq:surplus-partition}
W(N)=\sum_{i\in I}m_i+(|S|+1)\sigma.
\end{equation}
\begin{claim}$\pi^{LDV}\in\Core(N,W)$.\end{claim}
\begin{proof}
Since buyers are substitutes, the equivalence in \eqref{eq:substitutes-vcg-core} implies that
\[\pi^{VCG}=\Big(W(N)-\sum_{i\in B}m_i,(m_i)_{i\in B}\Big)\in\Core(N,W).\]
Starting from $\pi^{VCG}$, transfer $m_i-\sigma$ from each buyer $i\in S$ to the seller, leaving the payoffs of buyers in $I$ unchanged. The resulting payoff of each buyer is $\min\{\sigma,m_i\}$. The seller receives
\begin{align*}
W(N)-\sum_{i\in B}m_i+\sum_{i\in S}(m_i-\sigma)
&=W(N)-\sum_{i\in I}m_i-|S|\sigma\\
&=\sigma,
\end{align*}
where the last equality follows from \eqref{eq:surplus-partition}. Thus the resulting vector is $\pi^{LDV}$. Since each transfer belongs to $[0,m_i]$, so repeated application of the Transfer Lemma (Lemma~\ref{lem:transfer}) gives $\pi^{LDV}\in\Core(N,W)$.
\end{proof}
\begin{claim}\label{claim:ldv-dominates}$\pi^{LDV}$ Lorenz dominates every payoff vector in $\Core(N,W)$.\end{claim}
\begin{proof}
Fix $\pi\in\Core(N,W)$. If $W(N)=0$, then every component of both $\pi$ and $\pi^{LDV}$ is zero, so the conclusion is immediate. Suppose therefore that $W(N)>0$. In this case $\sigma>0$.

If $m_i=0$ for every buyer $i$, the Cap Lemma (Lemma~\ref{lem:cap}) and nonnegativity give $\pi_i=0$ for every buyer, so $\pi=(W(N),0,\dots,0)$. In this case $\sigma=W(N)$ and $\pi^{LDV}=\pi$, so the conclusion is immediate. Suppose therefore that $m_i>0$ for some buyer $i$. Then $\sum_{i\in B}\min\{\sigma,m_i\}>0$, and the equation defining $\sigma$ gives $\sigma<W(N)$. Thus
\[0<\sigma<W(N).\]
Buyers with a zero marginal contribution also need separate treatment. Let $I_0=\{i\in B:m_i=0\}$ and $I^+=I\setminus I_0$. Since $\sigma>0$, every buyer in $I_0$ belongs to $I$. For $i\in I_0$, the Cap Lemma and nonnegativity give $\pi_i=0=\pi_i^{LDV}$. For $i\in I^+$, we have $0<m_i<\sigma<W(N)$.

We have,
\[\pi_i^{LDV}=m_i\quad\text{for }i\in I,\qquad\pi_j^{LDV}=\sigma\quad\text{for }j\in S\cup\{s\}.\]
By the Cap Lemma (Lemma~\ref{lem:cap}), $\pi_i\leq m_i$ for every buyer $i$. Define
\begin{equation}\label{eq:delta}
\delta=\sum_{i\in I}(m_i-\pi_i)\geq0.
\end{equation}
Since both payoff vectors have total $W(N)$,
\begin{align*}
\sum_{j\in S\cup\{s\}}(\pi_j-\sigma)
&=W(N)-\sum_{i\in I}\pi_i-(|S|+1)\sigma\\
&=\sum_{i\in I}(m_i-\pi_i)=\delta\qquad\text{(by \eqref{eq:surplus-partition})}
\end{align*}
Fix an arbitrary continuous, increasing, and concave function $u:[0,W(N)]\to\mathbb R$. From Fact~\ref{fact:hardy-littlewood-polya} showing $\pi^{LDV}\succeq_L\pi$ is equivalent to showing
\[\sum_{j\in N}u(\pi_j^{LDV})\geq\sum_{j\in N}u(\pi_j).\]
We use a standard property of a concave function. At every interior point of its domain, a concave function has a supergradient. Thus, if $0<y<W(N)$, there exists a number $\lambda(y)$ such that
\begin{equation}\label{eq:supergradient}
u(x)\leq u(y)+\lambda(y)(x-y)\qquad\text{for every }x\in[0,W(N)].
\end{equation}
For each $i\in I^+$, apply \eqref{eq:supergradient} with $x=\pi_i$ and $y=m_i$; this is valid because $0<m_i<W(N)$. Summing gives
\begin{align*}
\sum_{i\in I^+}u(\pi_i)&\leq\sum_{i\in I^+}u(m_i)+\sum_{i\in I^+}\lambda(m_i)(\pi_i-m_i)\\
&\leq\sum_{i\in I^+}u(m_i)+\lambda(\sigma)\sum_{i\in I^+}(\pi_i-m_i)\\
&=\sum_{i\in I^+}u(m_i)-\lambda(\sigma)\delta\qquad\text{(by \eqref{eq:delta})}
\end{align*}
where, second inequality follows from the fact that $(\pi_i-m_i)\leq0$ (Cap Lemma), $m_i<\sigma$ for every buyer $i\in I^+$ and supergradient of a concave function is weakly decreasing. The last equality uses $m_i-\pi_i=0$ for every $i\in I_0$, so that $\delta=\sum_{i\in I^+}(m_i-\pi_i)$. Adding $u(\pi_i)=u(m_i)$ for every $i\in I_0$ gives
\[\sum_{i\in I}u(\pi_i)\leq\sum_{i\in I}u(m_i)-\lambda(\sigma)\delta.\]
For each $j\in S\cup\{s\}$, apply \eqref{eq:supergradient} with $x=\pi_j$ and $y=\sigma$; this is valid because $0<\sigma<W(N)$. Summing and using $\sum_{j\in S\cup\{s\}}(\pi_j-\sigma)=\delta$ gives
\[\sum_{j\in S\cup\{s\}}u(\pi_j)\leq(|S|+1)u(\sigma)+\lambda(\sigma)\delta.\]
Adding these inequalities yields
\begin{align*}
\sum_{j\in N}u(\pi_j)&\leq\sum_{i\in I}u(m_i)+(|S|+1)u(\sigma)\\
&=\sum_{j\in N}u(\pi_j^{LDV}).
\end{align*}
Since $u$ was arbitrary continuous, increasing and concave function, Fact~\ref{fact:hardy-littlewood-polya} implies that $\pi^{LDV}\succeq_L\pi$. Since $\pi$ was an arbitrary core payoff, the claim follows.
\end{proof}
\begin{claim}$\pi^{LDV}$ is the only Lorenz-dominant vector in $\Core(N,W)$.\end{claim}
\begin{proof}
If $W(N)=0$, the core contains only the zero vector, and there is nothing to prove. Suppose therefore that $W(N)>0$. Let $y\in\Core(N,W)$ be a Lorenz-dominant vector, so that $y\succeq_L\pi^{LDV}$. By Claim~\ref{claim:ldv-dominates}, $\pi^{LDV}\succeq_L y$. Mutual Lorenz dominance means that the cumulative sums of the ordered components coincide at every rank, so $y$ and $\pi^{LDV}$ have the same ordered components. In particular,
\begin{equation}\label{eq:same-ordered-sum}
\sum_{j\in N}u(y_j)=\sum_{j\in N}u(\pi^{LDV}_j)
\end{equation}
for every function $u$.

Suppose, to the contrary, that $y\neq\pi^{LDV}$. The core is convex, since it is defined by one linear equation and finitely many linear inequalities. Hence $z=\frac12\big(y+\pi^{LDV}\big)\in\Core(N,W)$. Let $u(x)=\sqrt{x}$ on $[0,W(N)]$, which is continuous, increasing and strictly concave. Strict concavity gives $u(z_j)\geq\frac12u(y_j)+\frac12u(\pi^{LDV}_j)$ for every $j$, with strict inequality whenever $y_j\neq\pi^{LDV}_j$. Since $y\neq\pi^{LDV}$, at least one inequality is strict. Summing and using \eqref{eq:same-ordered-sum},
\[
\sum_{j\in N}u(z_j)>\frac12\sum_{j\in N}u(y_j)+\frac12\sum_{j\in N}u(\pi^{LDV}_j)=\sum_{j\in N}u(\pi^{LDV}_j).
\]
But Claim~\ref{claim:ldv-dominates} gives $\pi^{LDV}\succeq_L z$, and Fact~\ref{fact:hardy-littlewood-polya} then gives $\sum_{j\in N}u(\pi^{LDV}_j)\geq\sum_{j\in N}u(z_j)$, a contradiction. Hence $y=\pi^{LDV}$.
\end{proof}
It remains to establish the second characterisation of the seller's payoff.
\begin{claim}
\[\sigma=\max_{T\subseteq B}\frac{W(N)-\sum_{i\in T}m_i}{|B\setminus T|+1}.\]
\end{claim}
\begin{proof}
For each $T\subseteq B$, define
\[h(T)=\frac{W(N)-\sum_{i\in T}m_i}{|B\setminus T|+1}.\]
Using the equation defining $\sigma$, together with $\min\{\sigma,m_i\}\leq m_i$ and $\min\{\sigma,m_i\}\leq\sigma$, gives
\begin{align*}
W(N)&=\sigma+\sum_{i\in T}\min\{\sigma,m_i\}+\sum_{i\in B\setminus T}\min\{\sigma,m_i\}\\
&\leq\sum_{i\in T}m_i+(|B\setminus T|+1)\sigma.
\end{align*}
Rearranging yields $h(T)\leq\sigma$ for every $T\subseteq B$.
For $T=I$, equation~\eqref{eq:surplus-partition} gives
\[h(I)=\frac{W(N)-\sum_{i\in I}m_i}{|S|+1}=\sigma.\]
Hence the maximum equals $\sigma$ and is attained at $T=I$.
\end{proof}
\end{proof}

\subsection{An alternative proof via a convex game}\label{app:existence-convex}
This subsection gives a second proof of Theorem~\ref{thm:ldv-existence}. It shows that, when buyers are substitutes, the core of $(N,W)$ coincides with the core of a convex game, and then applies \citet{dutta1989concept}. The argument also isolates where the substitutes condition enters.

Let
\begin{equation}\label{eq:vcg-seller-residual}
r=W(N)-\sum_{i\in B}m_i=\pi^{VCG}_s
\end{equation}
denote the seller's VCG payoff. Every $m_i\geq0$, because any allocation feasible for $B\setminus\{i\}$ is feasible for $B$. Applying the substitutes condition to $T=B$ gives $m_B=W(N)-W(\{s\})=W(N)\geq\sum_{i\in B}m_i$, so $r\geq0$.

\begin{lemma}[Box Lemma]\label{lem:box}
Suppose buyers are substitutes. Then $\Core(N,W)=K$, where
\[
K=\Big\{x\in\R^N_+:\sum_{j\in N}x_j=W(N)\text{ and }x_i\leq m_i\text{ for every }i\in B\Big\}.
\]
\end{lemma}
\begin{proof}
($\subseteq$) Let $x\in\Core(N,W)$. Every singleton coalition has value zero, since $W(\{s\})=U(\emptyset)=0$ and $W(\{i\})=0$ for $i\in B$. Hence $x_j\geq0$ for every $j$. The constraint for $N\setminus\{i\}$, together with efficiency, gives $x_i\leq W(N)-W(N\setminus\{i\})=m_i$. This inclusion does not use the substitutes condition.

($\supseteq$) Let $x\in K$ and $C\subseteq N$. If $s\notin C$, then $W(C)=0\leq x(C)$ because $x\geq0$. If $s\in C$, let $T=B\setminus C$, so that $C=N\setminus T$. Then
\[
x(C)=W(N)-\sum_{i\in T}x_i\;\geq\;W(N)-\sum_{i\in T}m_i\;\geq\;W(N)-m_T=W(N\setminus T)=W(C).
\]
The first inequality uses the caps. The second uses the substitutes condition. Hence $x\in\Core(N,W)$.
\end{proof}

In words, every coalitional constraint is implied by the single-buyer caps. The substitutes condition is used only in the second inclusion, and it cannot be dropped there. In Example~\ref{ex:vcg-core-failure}, $\pi^{VCG}=(0,0,2,2)$ lies in $K$ but violates the constraint of $\{s,1\}$.

Define the \emph{Vickrey game} $(N,W^V)$ by
\begin{equation}\label{eq:vickrey-game}
W^V(C)=
\begin{cases}
0, & \text{if } s\notin C,\\[2pt]
r+\displaystyle\sum_{i\in C\cap B}m_i=\sum_{j\in C}\pi^{VCG}_j, & \text{if } s\in C.
\end{cases}
\end{equation}
Each coalition containing the seller is worth the total VCG payoff of its members.

\begin{lemma}\label{lem:vickrey-game}
Suppose buyers are substitutes. Then
\begin{enumerate}[label=(\alph*)]
\item $W^V$ is convex: $W^V(C\cup D)+W^V(C\cap D)\geq W^V(C)+W^V(D)$ for all $C,D\subseteq N$;
\item $W^V(N)=W(N)$ and $\Core(N,W^V)=K=\Core(N,W)$.
\end{enumerate}
\end{lemma}
\begin{proof}
(a) For $A\subseteq N$, let $u_A$ be the unanimity game: $u_A(C)=1$ if $A\subseteq C$ and $u_A(C)=0$ otherwise. Then
\[
W^V=r\,u_{\{s\}}+\sum_{i\in B}m_i\,u_{\{s,i\}}.
\]
Each unanimity game is convex. If $A\subseteq C$ and $A\subseteq D$, both sides of the convexity inequality equal $2$. If $A$ lies in exactly one of $C$ and $D$, the right-hand side equals $1$, and the left-hand side is at least $1$ because $A\subseteq C\cup D$. Otherwise the right-hand side is $0$. Since $r\geq0$ and $m_i\geq0$, $W^V$ is a nonnegative combination of convex games and is therefore convex.

(b) First, $W^V(N)=r+\sum_{i\in B}m_i=W(N)$ by \eqref{eq:vcg-seller-residual}. Fix an efficient $x$. For a coalition $C$ with $s\in C$, let $T=B\setminus C$. Since $x(C)=W(N)-\sum_{i\in T}x_i$ and $W^V(C)=W(N)-\sum_{i\in T}m_i$, the core constraint of $C$ in $W^V$ reads
\begin{equation}\label{eq:vickrey-constraint}
\sum_{i\in T}x_i\leq\sum_{i\in T}m_i.
\end{equation}
For a coalition $C$ with $s\notin C$, the constraint reads $x(C)\geq0$.

Suppose $x\in K$. Then $x(C)\geq0$ for every $C$, and summing the caps over $T$ gives \eqref{eq:vickrey-constraint}. Hence $x\in\Core(N,W^V)$. Conversely, suppose $x\in\Core(N,W^V)$. The singleton constraints give $x_i\geq0$ for every buyer and $x_s\geq W^V(\{s\})=r\geq0$. Taking $T=\{i\}$ in \eqref{eq:vickrey-constraint} gives $x_i\leq m_i$. Hence $x\in K$. Lemma~\ref{lem:box} gives $K=\Core(N,W)$.
\end{proof}

Convexity of $W^V$ needs only $r\geq0$. The substitutes condition matters for the equality $\Core(N,W^V)=\Core(N,W)$.

We use the following result of \citet{dutta1989concept} for convex games.
\begin{fact}[\cite{dutta1989concept}]\label{fact:dutta-ray}
Let $(N,v)$ be a convex game with $v(\emptyset)=0$. Then $\Core(N,v)$ contains a unique allocation that Lorenz dominates every other core allocation, the egalitarian solution $E(v)$. It is computed as follows. Set $U_0=\emptyset$. At step $k\geq1$, among nonempty $E\subseteq N\setminus U_{k-1}$, let $C_k$ be the largest coalition that maximises
\[
\frac{v(E\cup U_{k-1})-v(U_{k-1})}{|E|},
\]
and give each member of $C_k$ this maximal value. Set $U_k=U_{k-1}\cup C_k$ and stop once $U_k=N$.
\end{fact}
Convexity guarantees that the union of maximisers is again a maximiser at each step, so the largest maximiser $C_k$ is well defined.

\begin{proof}[Alternative proof of Theorem~\ref{thm:ldv-existence}]
If $W(N)=0$, then $r=0$ and every $m_i=0$, so $K=\{0\}$ and $\sigma=0$. The claim is immediate. Suppose therefore that $W(N)>0$. Then $\sigma\geq W(N)/(|B|+1)>0$.

By Lemma~\ref{lem:vickrey-game}, $\Core(N,W)=\Core(N,W^V)$ and $W^V$ is convex. Lorenz dominance over a set depends only on the set. Fact~\ref{fact:dutta-ray} therefore shows that $\Core(N,W)$ contains a unique Lorenz-dominant vector, namely $E(W^V)$. It remains to show that $E(W^V)=\pi^{LDV}$.

\emph{Step 1.} A coalition without the seller has average value $0<\sigma$. For a coalition $\{s\}\cup D$ with $D\subseteq B$, write
\[
a(D)=\frac{W^V(\{s\}\cup D)}{|D|+1}=\frac{r+\sum_{i\in D}m_i}{|D|+1}.
\]
Substitute $r=W(N)-\sum_{i\in B}m_i$ and $W(N)=\sigma+\sum_{i\in B}\min\{\sigma,m_i\}$. Using $\min\{\sigma,m_i\}-\sigma=-(\sigma-m_i)^+$ and $\min\{\sigma,m_i\}-m_i=-(m_i-\sigma)^+$, we obtain
\begin{equation}\label{eq:average-gap}
(|D|+1)\big(a(D)-\sigma\big)=-\sum_{i\in D}(\sigma-m_i)^+\;-\sum_{i\in B\setminus D}(m_i-\sigma)^+.
\end{equation}
The right-hand side is nonpositive, so $a(D)\leq\sigma$. It equals zero if and only if $D$ contains every buyer with $m_i>\sigma$ and no buyer with $m_i<\sigma$. The maximal average is therefore $\sigma$, and the largest maximiser is $C_1=\{s\}\cup S$, where $S=\{i\in B:m_i\geq\sigma\}$. The seller and every buyer in $S$ receive $\sigma$.

\emph{Later steps.} Let $I=B\setminus S$. If $I=\emptyset$, the algorithm stops. Otherwise, for every $U$ with $C_1\subseteq U$ and every nonempty $E\subseteq N\setminus U$, both $E\cup U$ and $U$ contain the seller, so
\[
W^V(E\cup U)-W^V(U)=\sum_{i\in E}m_i.
\]
This marginal game is additive. At each later step, the maximal average is the largest $m_i$ among the remaining buyers, and the largest maximiser is the set of remaining buyers attaining it. Each of them receives her own $m_i$. Hence every buyer in $I$ receives $m_i$.

Therefore $E(W^V)_s=\sigma$, $E(W^V)_i=\sigma=\min\{\sigma,m_i\}$ for $i\in S$, and $E(W^V)_i=m_i=\min\{\sigma,m_i\}$ for $i\in I$. That is, $E(W^V)=\pi^{LDV}$.
\end{proof}

Writing $T=B\setminus D$, the average $a(D)$ equals $\big(W(N)-\sum_{i\in T}m_i\big)/(|B\setminus T|+1)$. Equation~\eqref{eq:average-gap} thus also recovers characterisation~(ii) of Theorem~\ref{thm:ldv-existence}: the maximum of this expression over $T\subseteq B$ is $\sigma$, attained at $T=I$.

The argument also explains why the substitutes condition cannot simply be dropped. Without it, $W^V$ may still be convex, but its core $K$ can strictly contain $\Core(N,W)$, and $E(W^V)$ can lie outside $\Core(N,W)$. In Example~\ref{ex:vcg-core-failure}, $r=0$, $(m_1,m_2,m_3)=(0,2,2)$ and $W(N)=4$, so $\sigma=4/3$ and $E(W^V)=(4/3,0,4/3,4/3)$. This vector gives coalition $\{s,1\}$ only $4/3<2=W(\{s,1\})$.

\section{Proof of Theorem \ref{thm:seller-proposer-implementation}}\label{app:implementation}
Subsection~\ref{subsec:equilibrium-proof-idea} constructs an equilibrium with immediate agreement at every node. We establish the properties used in the proof idea and then prove that every subgame-perfect equilibrium yields the target payoffs. The marginal contributions $m_i$ remain fixed throughout. The buyers-are-substitutes condition identifies the initial target vector with the Lorenz-dominant vector by Theorem~\ref{thm:ldv-existence}.
\begin{lemma}\label{lem:node-common-level}
For every finite set $C\subseteq B$ and every $w\geq0$, equation~\eqref{eq:node-common-level} has a unique nonnegative solution $\sigma_{C,w}$. Holding $C$ fixed, $\sigma_{C,w}$ is strictly increasing in $w$.
\end{lemma}
\begin{proof}
The left-hand side of \eqref{eq:node-common-level} is continuous and strictly increasing from zero to infinity, which gives existence and uniqueness. If $w'>w$, strict monotonicity of the same expression implies $\sigma_{C,w'}>\sigma_{C,w}$.
\end{proof}
The next lemma formalises the consistency property: when a buyer leaves with her target payoff, the target payoffs of all remaining players are unchanged.

\begin{lemma}
\label{lem:target-consistency}
Fix a node $(C,w)$ and a buyer $i\in C$. If buyer $i$ exits with
\[
    \ell_i(C,w)=\min\{m_i,\sigma_{C,w}\},
\]
then
\[
    \sigma_{C\setminus\{i\},\,w-\ell_i(C,w)}
    =
    \sigma_{C,w}.
\]
\end{lemma}

\begin{proof}
Let $\tau=\sigma_{C,w}$. By its defining equation,
\[
    \tau+\sum_{h\in C}\min\{m_h,\tau\}=w.
\]
Subtracting, $\min\{m_i,\tau\}=\ell_i(C,w)$ from both sides gives
\[
    \tau
    +
    \sum_{h\in C\setminus\{i\}}\min\{m_h,\tau\}
    =
    w-\ell_i(C,w).
\]
Thus, $\tau$ satisfies the defining equation for the common level at the
reduced node $ \left(C\setminus\{i\},w-\ell_i(C,w)\right).$
Uniqueness in Lemma~\ref{lem:node-common-level} therefore implies
\[
    \sigma_{C\setminus\{i\},\,w-\ell_i(C,w)}
    =
    \tau
    =
    \sigma_{C,w}.
\]
\end{proof}

\begin{lemma}
\label{lem:exit-payment-feasibility}
For every node $(C,w)$, every buyer $i\in C$, and every feasible seller
share $z\in[0,w]$,
\[
    0\leq e_i(C,z;w)\leq w.
\]
Consequently, the remaining surplus after an exit is nonnegative.
\end{lemma}

\begin{proof}
Every allocation feasible for the buyer set $B \setminus \{i\}$ remains feasible for $B$
(simply leave buyer $i$ empty-handed), so $W(N) \ge W(N \setminus \{i\})$, i.e.
\[
m_i = W(N) - W(N \setminus \{i\}) \ge 0 \qquad \forall i \in B.
\]
Since $z \ge 0$, it follows that
\[
0 \le A_i(z) = \min\{m_i, z\} \le z \le w.
\]
Similarly, $m_h \ge 0$ and $z \ge 0$ imply $\min\{m_h, z\} \ge 0$ for every $h \in C \setminus \{i\}$.
Hence
\[
R_i(C,z;w) = w - z - \sum_{h \in C \setminus \{i\}} \min\{m_h, z\} \le w - z \le w.
\]
Thus, both claims entering the maximum that defines $e_i(C,z;w)$ are at most $w$. Moreover,
the capped claim is nonnegative. Therefore,
\[
0 \le e_i(C,z;w) = \max\{A_i(z), R_i(C,z;w)\} \le w.
\]
It follows that $w - e_i(C,z;w) \ge 0$.
\end{proof}

\begin{lemma}
\label{lem:responder-floor}
Fix a node $(C,w)$ and a buyer $i \in C$, and let $\sigma = \sigma_{C,w}$. For every
feasible seller share $z \in [0,w]$,
\[
e_i(C,z;w) \ge \ell_i(C,w).
\]
Moreover, at $z = \sigma$,
\[
e_i(C,\sigma;w) = \ell_i(C,w).
\]
\end{lemma}

\begin{proof}
Let $\ell_i = \ell_i(C,w) = \min\{m_i,\sigma\}$. At $z = \sigma$,
\[
A_i(\sigma) = \min\{m_i,\sigma\} = \ell_i.
\]
By the defining equation for $\sigma$,
\[
R_i(C,\sigma;w) = w - \sigma - \sum_{h \in C \setminus \{i\}} \min\{m_h,\sigma\} = \min\{m_i,\sigma\} = \ell_i.
\]
Hence $e_i(C,\sigma;w) = \max\{A_i(\sigma), R_i(C,\sigma;w)\} = \ell_i$, proving the equality claim.

For an arbitrary feasible $z$: if $z \le \sigma$, Remark \ref{rem:claim-monotonicity} gives
$R_i(C,z;w) \ge R_i(C,\sigma;w) = \ell_i$, so $e_i(C,z;w) \ge \ell_i$. If $z \ge \sigma$,
Remark \ref{rem:claim-monotonicity} gives $A_i(z) \ge A_i(\sigma) = \ell_i$, so again $e_i(C,z;w) \ge \ell_i$.
This covers every feasible $z$.
\end{proof}
In simple terms, a buyer can always secure at least her target payoff by
rejecting the seller's proposal. When the seller proposes her target share,
every buyer's exit payment is exactly her target payoff.

\begin{lemma}
\label{lem:floored-exit-continuation}
Fix a node $(C,w)$ and a buyer $i\in C$. If buyer $i$ exits with a feasible
payment
\[
    q\in[\ell_i(C,w),w],
\]
then
\[
    \sigma_{C\setminus\{i\},\,w-q}
    \leq
    \sigma_{C,w}.
\]
Equality holds if and only if
\[
    q=\ell_i(C,w).
\]
\end{lemma}

\begin{proof}
Lemma~\ref{lem:target-consistency} gives
\[
    \sigma_{C\setminus\{i\},\,w-\ell_i(C,w)}
    =
    \sigma_{C,w}.
\]
Since $q\geq\ell_i(C,w)$,
\[
    w-q\leq w-\ell_i(C,w).
\]
By the strict monotonicity established in
Lemma~\ref{lem:node-common-level},
\[
    \sigma_{C\setminus\{i\},\,w-q}
    \leq
    \sigma_{C\setminus\{i\},\,w-\ell_i(C,w)}
    =
    \sigma_{C,w}.
\]
The inequality is an equality if and only if
\[
    w-q=w-\ell_i(C,w),
\]
which is equivalent to $q=\ell_i(C,w)$.
\end{proof}

\begin{lemma}
\label{lem:seller-target-guarantee}
Fix a node $(C,w)$. Suppose that, at every node with strictly fewer than $|C|$ active buyers, a subgame-perfect equilibrium exists and every such equilibrium yields the target payoffs in \eqref{eq:node-target-payoffs}. Then the
seller can guarantee $\sigma_{C,w}$ at $(C,w)$ by proposing
$P^{\mathrm{tar}}(C,w)$.
\end{lemma}

\begin{proof}
Let $\sigma=\sigma_{C,w}$. If every buyer accepts
$P^{\mathrm{tar}}(C,w)$, the seller receives $\sigma$ directly.
Suppose instead that buyer $j$ is the first buyer to reject. The seller's
proposed share is $z=\sigma$, so Lemma~\ref{lem:responder-floor} gives
\[
    e_j(C,\sigma;w)
    =
    \ell_j(C,w).
\]
The game therefore continues at
\[
    \left(
        C\setminus\{j\},
        w-\ell_j(C,w)
    \right).
\]
By Lemma~\ref{lem:target-consistency},
\[
    \sigma_{C\setminus\{j\},\,w-\ell_j(C,w)}
    =
    \sigma_{C,w}
    =
    \sigma.
\]
The reduced node has $|C|-1$ active buyers. The supposition therefore gives the seller $\sigma$ in every subgame-perfect equilibrium
of the continuation game.

This argument does not depend on which buyer rejects first. In particular,
it remains valid if several buyers are indifferent between accepting and
rejecting and regardless of how any such indifference is resolved. Thus,
the target proposal guarantees the seller $\sigma_{C,w}$.
\end{proof}

\begin{lemma}
\label{lem:seller-upper-bound}
Fix a node $(C,w)$. Suppose that, at every node with strictly fewer than $|C|$ active buyers, a subgame-perfect equilibrium exists and every such equilibrium yields the target payoffs in \eqref{eq:node-target-payoffs}. Under
sequentially rational buyer responses, every feasible seller proposal at
$(C,w)$ gives the seller at most $\sigma_{C,w}$.
\end{lemma}

\begin{proof}
Let $\sigma=\sigma_{C,w},
    \;
    \ell_i=\ell_i(C,w).$
    
\noindent Consider an arbitrary feasible proposal $P=\left(z,(x_i)_{i\in C}\right).$

\noindent Suppose first that every buyer accepts. Along this continuation, buyer $i$
receives $x_i$. If she rejects at her response node, she instead receives
$e_i(C,z;w)$. Sequential rationality and
Lemma~\ref{lem:responder-floor} therefore require
\[
    x_i
    \geq
    e_i(C,z;w)
    \geq
    \ell_i
    \qquad
    \text{for every }i\in C.
\]
Using feasibility of the proposal and the defining equation for $\sigma$,
we obtain
\begin{align*}
    z
    &=
    w-\sum_{i\in C}x_i\\
    &\leq
    w-\sum_{i\in C}\ell_i\\
    &=
    w-\sum_{i\in C}\min\{m_i,\sigma\}\\
    &=
    \sigma.
\end{align*}
Thus, an accepted proposal gives the seller at most $\sigma$.

Suppose instead that buyer $j$ is the first buyer to reject. She exits with
\[
    q_j=e_j(C,z;w)\geq\ell_j(C,w)
\]
by Lemma~\ref{lem:responder-floor}. By
Lemma~\ref{lem:exit-payment-feasibility}, $q_j\leq w$, so the reduced node
is well defined. It contains $|C|-1$ active buyers. The supposition therefore gives the seller the continuation payoff
\[
    \sigma_{C\setminus\{j\},\,w-q_j}.
\]
Lemma~\ref{lem:floored-exit-continuation} implies
\[
    \sigma_{C\setminus\{j\},\,w-q_j}
    \leq
    \sigma_{C,w}
    =
    \sigma.
\]
Thus, a proposal that leads to a rejection also gives the seller at most
$\sigma$. The two cases exhaust all possible outcomes of a proposal.
\end{proof}

\begin{proof}[\textbf{Proof of Theorem}~\ref{thm:seller-proposer-implementation}]
Existence at every node was established in Subsection~\ref{subsec:equilibrium-proof-idea}. We prove by induction that every subgame-perfect equilibrium at every node $(C,w)$ gives each active player her target payoff.

If $C=\varnothing$, the rules of $\Gamma^S$ give the seller the entire
remaining surplus $w$. Moreover, the defining equation for
$\sigma_{\varnothing,w}$ is simply
\[
    x=w,
\]
so
\[
    \sigma_{\varnothing,w}=w.
\]
There are no active buyers. Thus, the payoff assertion holds at every node with no
active buyers.

Now fix $k\geq 1$ and suppose that the payoff assertion holds at every node with
fewer than $k$ active buyers. Consider a node $(C,w)$ with $|C|=k$, and
let
\[
    \sigma=\sigma_{C,w},
    \qquad
    \ell_i=\ell_i(C,w)
    \quad\text{for }i\in C.
\]
Every rejection removes exactly one buyer, so every continuation following
a rejection has strictly fewer than $k$ active buyers. The induction
hypothesis therefore applies to every such continuation.

Consider an arbitrary subgame-perfect equilibrium at $(C,w)$. By
Lemma~\ref{lem:seller-target-guarantee}, the seller can guarantee
$\sigma$. Since her equilibrium proposal must be optimal, her equilibrium
payoff is at least $\sigma$. By Lemma~\ref{lem:seller-upper-bound}, no
proposal can give her more than $\sigma$. Hence, her equilibrium payoff is
exactly
\[
    \sigma=\sigma_{C,w}.
\]

Suppose first that the equilibrium proposal
\[
    P=\left(z,(x_i)_{i\in C}\right)
\]
is accepted by every buyer. The seller then receives $z$, so the preceding
conclusion gives
\[
    z=\sigma.
\]
At the response node of buyer $i$, rejecting would give her
$e_i(C,z;w)\geq\ell_i$ by Lemma~\ref{lem:responder-floor}, whereas accepting
along the equilibrium continuation gives her $x_i$. Sequential rationality
therefore requires
\[
    x_i\geq\ell_i
    \qquad
    \text{for every }i\in C.
\]
Feasibility and $z=\sigma$ give
\begin{align*}
    \sum_{i\in C}x_i
    &=
    w-z
    =
    w-\sigma
    =
    \sum_{i\in C}\min\{m_i,\sigma\}
    =
    \sum_{i\in C}\ell_i.
\end{align*}
Every term $x_i-\ell_i$ is nonnegative, and their sum is zero. Hence,
\[
    x_i-\ell_i=0
    \qquad
    \text{for every }i\in C.
\]
Thus, $x_i=\ell_i(C,w)$ for every active buyer.

Suppose instead that buyer $j$ is the first buyer to reject. She exits
with
\[
    q_j=e_j(C,z;w)\geq\ell_j(C,w)
\]
by Lemma~\ref{lem:responder-floor}. The reduced node is
\[
    \left(C\setminus\{j\},w-q_j\right).
\]
By the induction hypothesis, the seller's continuation payoff is
\[
    \sigma'
    =
    \sigma_{C\setminus\{j\},\,w-q_j}.
\]
This continuation payoff is also her equilibrium payoff at the original
node because the rejected proposal is discarded and the seller receives
no payment before continuation. Her equilibrium payoff has already been
shown to equal $\sigma$. Therefore,
\[
    \sigma'=\sigma.
\]
The equality condition in
Lemma~\ref{lem:floored-exit-continuation} now implies
\[
    q_j=\ell_j(C,w).
\]
Buyer $j$ therefore receives her target payoff.

\noindent By Lemma~\ref{lem:target-consistency},
\[
    \sigma_{C\setminus\{j\},\,w-\ell_j(C,w)}
    =
    \sigma_{C,w}
    =
    \sigma.
\]
Consequently, for every remaining buyer $i\in C\setminus\{j\}$, her target
at the reduced node is
\[
    \min\left\{
        m_i,
        \sigma_{C\setminus\{j\},\,w-\ell_j(C,w)}
    \right\}
    =
    \min\{m_i,\sigma\}
    =
    \ell_i(C,w).
\]
The induction hypothesis gives each remaining buyer this payoff in the
continuation game. Thus, the rejecting buyer and every buyer who remains
active receive their targets at $(C,w)$.

The accepted and rejected cases exhaust all equilibrium paths. Hence every
subgame-perfect equilibrium at $(C,w)$ gives the seller and all active
buyers their target payoffs. This completes the induction establishing
payoff uniqueness.

Finally, at the initial node $(B,W(N))$,
\[
    \sigma_{B,W(N)}=\sigma,
    \qquad
    \ell_i(B,W(N))=\min\{m_i,\sigma\}.
\]
By Theorem~\ref{thm:ldv-existence}, this is precisely
$\pi^{\mathrm{LDV}}$. Hence $\Gamma^S$ fully implements the
Lorenz-dominant vector in subgame-perfect equilibrium.
\end{proof}

\begin{proof}[\textbf{Further details on Remark~\ref{rem:multiple-equilibria}}]
\label{rem:multiple-optimal-proposals}
More generally, any feasible division of the buyers' surplus can be optimal when the seller proposes her target share. At any node
$(C,w)$, consider any feasible proposal
\[
    \left(
        \sigma_{C,w},
        (x_i)_{i\in C}
    \right)
    \qquad\text{such that}\qquad
    x_i\geq 0
    \quad\text{and}\quad
    \sum_{i\in C}x_i=w-\sigma_{C,w}.
\]
Thus, the seller proposes her target share but may divide the remaining
surplus among the buyers in any way.

If every buyer accepts, sequential rationality requires
\[
    x_i\geq \ell_i(C,w)
    \qquad\text{for every }i\in C.
\]
Since
\[
    \sum_{i\in C}x_i
    =
    w-\sigma_{C,w}
    =
    \sum_{i\in C}\ell_i(C,w),
\]
these inequalities can all hold only if
\[
    x_i=\ell_i(C,w)
    \qquad\text{for every }i\in C.
\]
Hence, a proposal with a different division among the buyers cannot be
accepted by everyone.

If some buyer rejects, however, the seller's proposed share is still
$\sigma_{C,w}$. Lemma~\ref{lem:responder-floor} then gives the rejecting
buyer exactly her target payoff, and Lemma~\ref{lem:target-consistency}
preserves the target level in the reduced game. The continuation therefore
gives the seller and every remaining buyer their target payoffs. Thus, every
proposal of this kind gives the seller $\sigma_{C,w}$ and
ultimately implements the target vector, whether it is accepted or rejected.

Since Lemma~\ref{lem:seller-upper-bound} shows that the seller cannot obtain
more than $\sigma_{C,w}$, all these proposals are optimal. Consequently, the
game need not have a unique subgame-perfect equilibrium. Whenever more
than one division of the buyers' surplus is feasible, different optimal
proposals, together with different responses by indifferent buyers, generate
multiple subgame-perfect equilibria. Theorem~\ref{thm:seller-proposer-implementation}
states that their strategies may differ, while their final payoff vector is
always the same.
\end{proof}

\section{The Sequential Pooling Algorithm}\label{app:pooling}

\noindent\textbf{Sequential Pooling Algorithm (SPA)}

\noindent Initialize $\pi^0 = \pi^{VCG}$. At each stage $t = 0,1,2,\dots$, given the current
payoff vector $\pi^t$, let
\[
h^t = \max_{i \in B} \pi_i^t, \qquad G^t = \{i \in B : \pi_i^t = h^t\}
\]
where $h^t$ denotes the highest payoff currently held by any buyer at stage $t$ and $G^t$ denotes the set of all buyers who have payoff equal to $h^t$ at stage $t$. Hence $G^t$ denotes the richest group of buyers at stage $t$.

\begin{itemize}
\item[(i)] If $\pi_s^t \ge h^t$, i.e, the seller's payoff is already at least as large
as the richest group's payoff: stop and output
$\pi^t$.

\item[(ii)] Otherwise, compute the payoff level
\[
\bar h^t = \frac{\pi_s^t + |G^t|\, h^t}{|G^t|+1}
\]
where $\bar h^t$ denotes the payoff that the seller and the richest group would have if they were pooled together and split the sum of their current payoffs evenly. 

If every buyer belongs to the richest group, that is $G^t = B$, pool the seller with $G^t$: set
$\pi_s^{t+1} = \pi_i^{t+1} = \bar h^t$ for every $i \in G^t$, output $\pi^{t+1}$, and stop.

If instead $G^t \ne B$, let $\ell^t = \max_{i \in B \setminus G^t} \pi_i^t$
denote the payoff level of the next richest group of buyers.
\begin{itemize}
\item If $\bar h^t \ge \ell^t$, pooling the seller with $G^t$ does not overtake
the next group, pool the seller with $G^t$ : set
$\pi_s^{t+1} = \pi_i^{t+1} = \bar h^t$ for every $i \in G^t$, output $\pi^{t+1}$, and stop.
\item If $\bar h^t < \ell^t$, pooling would overtake the next tier, so instead
lower every buyer in $G^t$ to the level $\ell^t$ and transfer the resulting
surplus to the seller: set $\pi_i^{t+1} = \ell^t$ for every $i \in G^t$, set
$\pi_s^{t+1} = \pi_s^t + |G^t|(h^t-\ell^t)$, leave every other payoff unchanged,
and proceed to stage $t+1$.
\end{itemize}
\end{itemize}
Each step of type (ii) that does not stop merges the richest group with the group
immediately below it into a single group, so the number of distinct payoff levels
among the buyers strictly decreases at every such step. Since there are at most
$|B|$ levels initially, the algorithm stops after at most $|B|$ steps.

\begin{proposition}
\label{prop:algorithm-finds-ldv}
Suppose buyers are substitutes. Then the output of the Sequential Pooling Algorithm (SPA) equals $\pi^{LDV}$.
\end{proposition}

\begin{proof}
We first note a structural fact about the algorithm. Suppose stage $t$ executes the non-terminal branch of step (ii), producing $\pi^{t+1}$. This
branch requires $\pi_s^t + |G^t| h^t < (|G^t|+1)\ell^t$, which by definition of $\pi_s^{t+1}$ gives
\[
\pi_s^{t+1} = \pi_s^t + |G^t|(h^t - \ell^t) < \ell^t = h^{t+1}.
\]
So condition (i) fails at stage $t+1$. Hence, if the algorithm ever executes a
non-terminal branch of step (ii), condition (i) can never fire afterward: it can only fire, if
at all, at the very first stage $t=0$, before any transfer has been made.
Therefore, exactly one of the following two cases occur:

\smallskip
\noindent\textbf{Case 1:} Condition (i) fires at $t=0$.

\noindent No buyer's payoff is ever changed, so $\pi^T = \pi^{VCG}$. The stopping condition gives
$\pi_s^{VCG} = \pi_s^0 \ge h^0 = \max_{i \in B} m_i$, so by Corollary~\ref{cor:vcg-ldv},
$\pi^{VCG} = \pi^{LDV}$. Hence $\pi^T = \pi^{LDV}$.

\smallskip
\noindent\textbf{Case 2:} The algorithm terminates via a pooling step.

\noindent Let
$G = G^{T-1}$ be the richest group at the terminal stage. By construction
$\pi_i^T = \pi_s^T$ for every $i \in G$, and $\pi_j^T = m_j$ for every buyer
$j \notin G$, since a buyer's payoff changes only if she belongs to the
currently active richest group.

\noindent For a buyer $j \notin G$, if any (there are none if $G=B$): no transfer has
been made from buyer $j$ at any stage, since she never belonged to the richest
group at any stage up to and including $T-1$, so $\pi_j^T = m_j$. In
particular, $m_j \le \ell^{T-1}$, and the terminal stopping condition gives
$\ell^{T-1} \le \bar h^{T-1} = \pi_s^T$. Hence $\min\{\pi_s^T, m_j\} = m_j =
\pi_j^T$.

\noindent For a buyer $i \in G$: since $\pi_s^T = \bar h^{T-1}$ is a weighted average of
$\pi_s^{T-1} < h^{T-1}$ and $h^{T-1}$, we have $\pi_s^T \le h^{T-1} \le m_i$,
the last inequality because a buyer's payoff is non-increasing while she
remains in the richest group, so her original marginal contribution $m_i$ is
at least her value $h^{T-1}$ just before the final pooling step. Hence
$\min\{\pi_s^T, m_i\} = \pi_s^T = \pi_i^T$.

\noindent So $\pi_j^T = \min\{\pi_s^T, m_j\}$ for every buyer $j \in B$ in this case.
Since the algorithm only moves payoff between a buyer and the seller, the
total is conserved, $\sum_{j \in N} \pi_j^T = \sum_{j \in N} \pi_j^{VCG} =
W(N)$, so
\[
\pi_s^T + \sum_{j \in B} \min\{\pi_s^T, m_j\} = W(N).
\]
Thus $\pi_s^T$ solves the same equation that uniquely characterizes $\sigma$
in the proof of Theorem~\ref{thm:ldv-existence}, so by uniqueness $\pi_s^T =
\sigma$. Hence $\pi_j^T = \min\{\sigma, m_j\}$ for every buyer $j$ and
$\pi_s^T = \sigma$, so $\pi^T = \pi^{LDV}$.
\end{proof}

The substitutes condition is needed only to identify the output with the Lorenz-dominant vector. The algorithm is well defined and terminates for every game, and its output always has the form $(\tau,(\min\{\tau,m_i\})_{i\in B})$, where $\tau$ solves $x+\sum_{i\in B}\min\{x,m_i\}=W(N)$. Without the substitutes condition, this vector need not lie in the core. In Example~\ref{ex:vcg-core-failure}, the algorithm returns $(4/3,0,4/3,4/3)$, which gives coalition $\{s,1\}$ only $4/3<2=W(\{s,1\})$, whereas the Lorenz-dominant core payoff is $(2,0,1,1)$ (Appendix~\ref{app:examples}).

\section{The substitutes condition: reasoning and proofs for the examples}\label{app:examples}
This appendix provides the detailed reasoning and proofs for the examples in Subsection~\ref{subsec:converse}.
\subsection{Notation and the Lorenz-maximality criterion}

We will use a result of \cite{hougaard2001set} to
identify Lorenz-maximal points of the core.  Since the notation needed for
their result is quite detailed, we introduce it step by step.  To make each
object concrete, we will use the auction from Example \ref{ex:vcg-core-failure} alongside.

In that example, there were three buyers and two goods.  The surplus generated
by each group of buyers is
\[
\begin{array}{c|cccccccc}
S
    & \varnothing & \{1\} & \{2\} & \{3\}
    & \{1,2\} & \{1,3\} & \{2,3\} & B \\
\hline
U(S)
    & 0 & 2 & 2 & 2 & 2 & 2 & 4 & 4
\end{array}
\]
Thus, for the induced game with player set $N=\{s,1,2,3\}$,
\[
    W(C)=
    \begin{cases}
        U(C\cap B), & \text{if }s\in C,\\
        0,           & \text{if }s\notin C.
    \end{cases}
\]
The core is
\begin{equation}
\label{eq:core-example-23}
\operatorname{Core}(N,W)
=
\left\{
x\in\mathbb R_+^4:
x_s+x_1+x_2+x_3=4,\quad
x_1=0,\quad
x_2+x_3\leq2
\right\}.
\end{equation}
We keep the core payoff
\[
    x^*=(2,0,1,1)
\]
in mind while introducing the notation.  Coordinates are ordered as
$(s,1,2,3)$.

For every coalition $C\subseteq N$, let $\mathbf 1_C\in\{0,1\}^{N}$
denote its incidence vector: its $j$th coordinate is one if $j\in C$ and
zero otherwise.  For instance, in Example \ref{ex:vcg-core-failure},
\[
    \mathbf 1_{\{s,2,3\}}=(1,0,1,1).
\]
For a core payoff $x$, define
\begin{equation}
\label{eq:def-Z}
    Z_W(x)
    =
    \left\{
       \mathbf 1_C:
       \varnothing\neq C\subset N,
       \ x(C)=W(C)
    \right\},
\end{equation}
where $x(C)=\sum_{j\in C}x_j$.  Thus, $Z_W(x)$ contains the incidence
vectors of the proper coalitions whose core constraints bind at $x$.

For $x^*=(2,0,1,1)$, the binding proper coalitions are $\{1\}$, $\{s,1\}$ and $\{s,2,3\}$, since 
\begin{align*}
    x^*(\{1\}) &= 0 = W(\{1\}), \\
    x^*(\{s,1\}) &= 2 = W(\{s,1\}), \\
    x^*(\{s,2,3\}) &= 4 = W(\{s,2,3\})
\end{align*}
No other proper coalition is tight.  Hence
\begin{equation}
\label{eq:Z-example-23}
    Z_W(x^*)
    =
    \left\{
       \mathbf 1_{\{1\}},
       \mathbf 1_{\{s,1\}},
       \mathbf 1_{\{s,2,3\}}
    \right\}.
\end{equation}

We next introduce the orderings used by \cite{hougaard2001set}.  Let $\mathcal O$ be the set of all orderings of the
players in $N$.  We write an ordering as
\[
    \rho=(\rho(1),\ldots,\rho(|N|)),
\]
where $\rho(1)$ is the first player in the ordering and $\rho(|N|)$ is the
last.  For a payoff vector $x$, let
\begin{equation}
\label{eq:def-Ox}
    \mathcal O(x)
    =
    \left\{
       \rho\in\mathcal O:
       x_{\rho(1)}\leq x_{\rho(2)}\leq\cdots\leq x_{\rho(|N|)}
    \right\}.
\end{equation}
Thus, $\mathcal O(x)$ contains all orderings from the poorest player to the
richest player that are consistent with $x$.  When players are tied, every
ordering of the tied players is included. In Example \ref{ex:vcg-core-failure}, $\mathcal O$ contains all $4!$ orderings of
$\{s,1,2,3\}$.  At $x^*$,
\[
    x_1^*<x_2^*=x_3^*<x_s^*.
\]
The only choice is therefore the order of buyers $2$ and $3$, and
\begin{equation}
\label{eq:O-example-23}
    \mathcal O(x^*)
    =
    \bigl\{(1,2,3,s),(1,3,2,s)\bigr\}.
\end{equation}

Each ordering $\rho\in\mathcal O$ generates a collection $P_\rho$ of
incidence vectors.  Starting with the first player in the ordering, we add
one player at a time and stop just before reaching the grand coalition:
\begin{equation}
\label{eq:def-Prho}
    P_\rho
    =
    \left\{
       \mathbf 1_{\{\rho(1)\}},
       \mathbf 1_{\{\rho(1),\rho(2)\}},
       \ldots,
       \mathbf 1_{\{\rho(1),\ldots,\rho(|N|-1)\}}
    \right\}.
\end{equation}
For the two orderings in \eqref{eq:O-example-23}, these collections are
\[
\begin{aligned}
P_{(1,2,3,s)}
    &=
    \left\{
       \mathbf 1_{\{1\}},
       \mathbf 1_{\{1,2\}},
       \mathbf 1_{\{1,2,3\}}
    \right\},\\
P_{(1,3,2,s)}
    &=
    \left\{
       \mathbf 1_{\{1\}},
       \mathbf 1_{\{1,3\}},
       \mathbf 1_{\{1,2,3\}}
    \right\}.
\end{aligned}
\]
Finally, define
\begin{equation}
\label{eq:def-Px}
    P(x)=\bigcup_{\rho\in\mathcal O(x)}P_\rho.
\end{equation}
Thus, $P(x)$ contains every poorest-player group obtained from an ordering
consistent with $x$.  In Example \ref{ex:vcg-core-failure},
\begin{equation}
\label{eq:P-example-23}
    P(x^*)
    =
    \left\{
       \mathbf 1_{\{1\}},
       \mathbf 1_{\{1,2\}},
       \mathbf 1_{\{1,3\}},
       \mathbf 1_{\{1,2,3\}}
    \right\}.
\end{equation}
A finite collection $D$ of vectors in $\mathbb R^N$ is called
\emph{balanced} if there are strictly positive numbers
$\lambda_d>0$ for every $d\in D$ such that
\begin{equation}
\label{eq:def-balanced-vectors}
    \sum_{d\in D}\lambda_d d=\mathbf 1_N.
\end{equation}
In words, the vectors in $D$ can be given positive weights so that every
player receives total weight one. We can now state the characterization.

\begin{fact}[\cite{hougaard2001set}]
\label{fact:hougaard-lorenz-maximal}
Let $(N,W)$ be a balanced game, and let
$x\in\operatorname{Core}(N,W)$.  Then $x$ is Lorenz maximal in the core if and only if $Z_W(x)\cup P(x)$ is balanced.
\end{fact}

The two parts of the collection play different roles.  The vectors in
$Z_W(x)$ record the core constraints that bind at $x$, while the vectors in
$P(x)$ record all payoff orderings relevant for the Lorenz comparison.  Fact \ref{fact:hougaard-lorenz-maximal} says that these two pieces of information together provide a complete
test for Lorenz maximality.

\subsection{An LDV exists without the substitutes condition}

We first apply the theorem to $x^*=(2,0,1,1)$ in Example \ref{ex:vcg-core-failure}.  Combining
\eqref{eq:Z-example-23} and \eqref{eq:P-example-23} gives
\[
\begin{aligned}
Z_W(x^*)\cup P(x^*)
=\bigl\{&
\mathbf 1_{\{1\}},
\mathbf 1_{\{s,1\}},
\mathbf 1_{\{s,2,3\}},
\mathbf 1_{\{1,2\}},
\mathbf 1_{\{1,3\}},
\mathbf 1_{\{1,2,3\}}
\bigr\}
\end{aligned}
\]
This collection is balanced, since
\begin{align}
\mathbf 1_N={}&
 \frac{1}{10}\mathbf 1_{\{1\}}
 +\frac{4}{10}\mathbf 1_{\{s,1\}}
 +\frac{6}{10}\mathbf 1_{\{s,2,3\}}
 +\frac{1}{10}\mathbf 1_{\{1,2\}}
 +\frac{1}{10}\mathbf 1_{\{1,3\}}
 +\frac{3}{10}\mathbf 1_{\{1,2,3\}}
\label{eq:balance-example-23}
\end{align}
Every coefficient is strictly positive. Fact \ref{fact:hougaard-lorenz-maximal}
therefore implies that $x^*$ is Lorenz maximal in the core. For this example we can say more: $x^*$ Lorenz dominates every core payoff.  Take any
$x\in\operatorname{Core}(N,W)$, and write its components in increasing
order as $x_{(1)}\leq x_{(2)}\leq x_{(3)}\leq x_{(4)}.$
By \eqref{eq:core-example-23}, every core payoff satisfies
\[
    x_1=0,
    \qquad
    x_2+x_3\leq2,
    \qquad
    x_s+x_1+x_2+x_3=4.
\]
Since all core payoffs are nonnegative and $x_1=0$, the smallest component
of $x$ is zero.  Hence $x_{(1)}=0.$ Next, because $x_2+x_3\leq2$, we have
\[
    \min\{x_2,x_3\}
    \leq \frac{x_2+x_3}{2}
    \leq1.
\]
Moreover,
\[
    x_s=4-x_2-x_3\geq2.
\]
Therefore, the second-smallest component of $x$ cannot exceed
$\min\{x_2,x_3\}$.  Since the smallest component is $x_1=0$, it follows
that
\[
    x_{(1)}+x_{(2)}
    \leq 0+\min\{x_2,x_3\}
    \leq1.
\]
Finally, $x_s\geq2$, so the largest component of $x$ satisfies
\[
    x_{(4)}\geq x_s\geq2.
\]
Using efficiency,
\[
    x_{(1)}+x_{(2)}+x_{(3)}
    =
    4-x_{(4)}
    \leq2.
\]
The components of $x^*=(2,0,1,1)$, arranged in increasing order, are $(0,1,1,2),$
and hence its cumulative sums are $(0,1,2,4)$.
The preceding inequalities show that, for every
$x\in\operatorname{Core}(N,W)$,
\[
\begin{aligned}
    x_{(1)}
        &= 0,\\
    x_{(1)}+x_{(2)}
        &\leq 1,\\
    x_{(1)}+x_{(2)}+x_{(3)}
        &\leq 2,\\
    x_{(1)}+x_{(2)}+x_{(3)}+x_{(4)}
        &=4.
\end{aligned}
\]
 Thus, at every rank, the cumulative payoff of the poorest
players under $x^*$ is at least as large as under $x$.  Therefore, $x^*\succeq_L x
    \;
    \text{for every }x\in\operatorname{Core}(N,W).$ Hence $x^*$ is the Lorenz-dominant payoff in the core.

Buyers are not substitutes in this example. Their marginal contributions are $m_1=0$ and $m_2=m_3=2$. For $T=\{2,3\}$,
\[m_T=W(N)-W(N\setminus T)=U(B)-U(\{1\})=4-2=2,\]
whereas $m_2+m_3=4$. Thus $m_T<m_2+m_3$, violating Definition~\ref{def:substitutes}. The substitutes condition is therefore not necessary for an LDV to exist.
\subsection{No LDV without the substitutes condition}
Consider the coalitional values in Example~\ref{ex:no-substitutes-no-ldv}. The buyers' marginal contributions are $m_1=12-9=3$, $m_2=12-11=1$, and $m_3=12-7=5$. For $T=\{1,2\}$,
\[m_T=W(N)-W(N\setminus T)=U(B)-U(\{3\})=12-9=3,\]
whereas $m_1+m_2=4$. Hence $m_T<m_1+m_2$, so the buyers-are-substitutes condition fails in this example too.

The core constraints in this example can be written as
\begin{align*}
 x_1&\leq3, & x_2&\leq1, & x_3&\leq5,\\
 x_1+x_2&\leq3, & x_1+x_3&\leq11,
 & x_2+x_3&\leq5,
\end{align*}
together with $x\geq0$ and
\[
    x_s+x_1+x_2+x_3=12.
\]
Consider
\[
    x^A=\left(\frac92,\frac52,\frac12,\frac92\right)
    \qquad\text{and}\qquad
    x^B=(5,2,1,4).
\]
Both lie in the core. We now verify that both are Lorenz-maximal using Fact \ref{fact:hougaard-lorenz-maximal}.

\noindent For $x^A$, the proper binding coalitions are $\{s,1\}$ and $\{s,3\}$.
Therefore
\[
    Z_W(x^A)
    =
    \left\{
       \mathbf 1_{\{s,1\}},
       \mathbf 1_{\{s,3\}}
    \right\}.
\]
Its payoff ranking is
\[
    x_2^A<x_1^A<x_s^A=x_3^A,
\]
so
\[
    \mathcal O(x^A)
    =
    \bigl\{(2,1,s,3),(2,1,3,s)\bigr\}.
\]
The two associated chains are
\[
\begin{aligned}
P_{(2,1,s,3)}
    &=
    \left\{
       \mathbf 1_{\{2\}},
       \mathbf 1_{\{1,2\}},
       \mathbf 1_{\{s,1,2\}}
    \right\},\\
P_{(2,1,3,s)}
    &=
    \left\{
       \mathbf 1_{\{2\}},
       \mathbf 1_{\{1,2\}},
       \mathbf 1_{\{1,2,3\}}
    \right\}.
\end{aligned}
\]
Hence
\[
    P(x^A)
    =
    \left\{
       \mathbf 1_{\{2\}},
       \mathbf 1_{\{1,2\}},
       \mathbf 1_{\{s,1,2\}},
       \mathbf 1_{\{1,2,3\}}
    \right\}.
\]
The collection $Z_W(x^A)\cup P(x^A)$ is balanced because
\begin{align*}
\mathbf 1_N={}&
 \frac15\mathbf 1_{\{s,1\}}
 +\frac35\mathbf 1_{\{s,3\}}
 +\frac15\mathbf 1_{\{2\}}
 +\frac15\mathbf 1_{\{1,2\}}
 +\frac15\mathbf 1_{\{s,1,2\}}
 +\frac25\mathbf 1_{\{1,2,3\}}
\end{align*}
All coefficients are strictly positive.  Thus,
Fact \ref{fact:hougaard-lorenz-maximal} implies that $x^A$ is Lorenz-maximal in the core.

\noindent For $x^B$, the proper binding coalitions are
$\{s,1\}$, $\{s,3\}$, and $\{s,1,3\}$.  Thus,
\[
    Z_W(x^B)
    =
    \left\{
       \mathbf 1_{\{s,1\}},
       \mathbf 1_{\{s,3\}},
       \mathbf 1_{\{s,1,3\}}
    \right\}.
\]
Its payoff ranking is
\[
    x_2^B<x_1^B<x_3^B<x_s^B.
\]
Consequently,
\[
    \mathcal O(x^B)=\bigl\{(2,1,3,s)\bigr\},
\]
and
\[
\begin{aligned}
P(x^B)=P_{(2,1,3,s)}
    =
    \left\{
       \mathbf 1_{\{2\}},
       \mathbf 1_{\{1,2\}},
       \mathbf 1_{\{1,2,3\}}
    \right\}.
\end{aligned}
\]
The collection $Z_W(x^B)\cup P(x^B)$ is balanced because
\begin{align*}
\mathbf 1_N={}&
 \frac15\mathbf 1_{\{s,1\}}
 +\frac25\mathbf 1_{\{s,3\}}
 +\frac25\mathbf 1_{\{s,1,3\}}
 +\frac35\mathbf 1_{\{2\}}
 +\frac15\mathbf 1_{\{1,2\}}
 +\frac15\mathbf 1_{\{1,2,3\}}
\end{align*}
Again, every coefficient is strictly positive.  Therefore $x^B$ is also
Lorenz maximal by Fact \ref{fact:hougaard-lorenz-maximal}. It remains to compare the two vectors.  Their ordered components and
cumulative sums are
\[
\begin{array}{c|c|c}
 & \text{ordered components} & \text{cumulative sums}\\
\hline
x^A
 & \left(\frac12,\frac52,\frac92,\frac92\right)
 & \left(\frac12,3,\frac{15}{2},12\right)\\[2mm]
x^B
 & (1,2,4,5)
 & (1,3,7,12).
\end{array}
\]
The payoff $x^B$ gives more to the poorest participant, since
$1>\frac12$.  In contrast, $x^A$ gives more in total to the three poorest
participants, since $\frac{15}{2}>7$.  Hence $x^A$ and $x^B$ are Lorenz
incomparable. If a Lorenz-dominant core payoff existed, it would Lorenz dominate both
$x^A$ and $x^B$ which is impossible because both are Lorenz-maximal.  Therefore this core contains no
Lorenz-dominant payoff.


\begin{thebibliography}{24}
\bibitem[Ausubel and Baranov(2026)]{ausubel2026vcg}
Ausubel, L.~M. and Baranov, O. (2026). The VCG mechanism, the core, and assignment stages in auctions.
\emph{Journal of Economic Theory}, 235:106192.

\bibitem[Ausubel and Milgrom(2002)]{ausubel2002ascending}
Ausubel, L. and Milgrom, P. (2002). Ascending auctions with package bidding.
\emph{Frontiers of theoretical economics}, 1(1):1019.

\bibitem[Ausubel et~al.(2006)]{ausubel2006lovely}
Ausubel, L.~M., Milgrom, P., et~al. (2006). The lovely but lonely vickrey auction.
\emph{Combinatorial auctions}, 17(3):22--26.

\bibitem[Bikhchandani and Mishra(2026)]{bikhchandani2026strategy}
Bikhchandani, S. and Mishra, D. (2026). Strategy-proof and efficient job matching with participation constraints.
\emph{arXiv preprint arXiv:2605.01715}.

\bibitem[Bikhchandani and Ostroy(2002)]{bikhchandani2002package}
Bikhchandani, S. and Ostroy, J.~M. (2002). The package assignment model.
\emph{Journal of Economic theory}, 107(2):377--406.

\bibitem[Bogomolnaia and Moulin(2004)]{bogomolnaia2004random}
Bogomolnaia, A. and Moulin, H. (2004). Random matching under dichotomous preferences.
\emph{Econometrica}, 72(1):257--279.

\bibitem[Bosmans and Lauwers(2011)]{bosmans2011lorenz}
Bosmans, K. and Lauwers, L. (2011). Lorenz comparisons of nine rules for the adjudication of conflicting claims.
\emph{International Journal of Game Theory}, 40(4):791--807.

\bibitem[Chatterjee et~al.(1993)]{chatterjee1993noncooperative}
Chatterjee, K., Dutta, B., Ray, D., and Sengupta, K. (1993). A noncooperative theory of coalitional bargaining.
\emph{The Review of Economic Studies}, 60(2):463--477.

\bibitem[Day and Milgrom(2008)]{day2008core}
Day, R. and Milgrom, P. (2008). Core-selecting package auctions.
\emph{international Journal of game Theory}, 36(3):393--407.

\bibitem[Day and Cramton(2012)]{day2012quadratic}
Day, R.~W. and Cramton, P. (2012). Quadratic core-selecting payment rules for combinatorial auctions.
\emph{Operations Research}, 60(3):588--603.

\bibitem[Dutta and Ray(1989)]{dutta1989concept}
Dutta, B. and Ray, D. (1989). A concept of egalitarianism under participation constraints.
\emph{Econometrica: Journal of the Econometric Society}, pages 615--635.

\bibitem[Gul and Stacchetti(1999)]{gul1999walrasian}
Gul, F. and Stacchetti, E. (1999). Walrasian equilibrium with gross substitutes.
\emph{Journal of Economic theory}, 87(1):95--124.

\bibitem[Hardy(1929)]{hardy1929some}
Hardy, G.~H. (1929). Some simple inequalities satisfied by convex functions.
\emph{Messenger Math.}, 58:145--152.

\bibitem[Holmstr\"om(1979)]{holmstrom1979groves}
Holmstr\"om, B. (1979). Groves' scheme on restricted domains.
\emph{Econometrica: Journal of the Econometric Society}, pages 1137--1144.

\bibitem[Hougaard et~al.(2001)]{hougaard2001set}
Hougaard, J.~L., Peleg, B., and Thorlund-Petersen, L. (2001). On the set of lorenz-maximal imputations in the core of a balanced game.
\emph{International Journal of Game Theory}, 30(2):147--165.

\bibitem[Ju and Wettstein(2009)]{ju2009implementing}
Ju, Y. and Wettstein, D. (2009). Implementing cooperative solution concepts: a generalized bidding approach.
\emph{Economic Theory}, 39(2):307--330.

\bibitem[Kelso Jr and Crawford(1982)]{kelso1982job}
Kelso Jr, A.~S. and Crawford, V.~P. (1982). Job matching, coalition formation, and gross substitutes.
\emph{Econometrica: Journal of the Econometric Society}, pages 1483--1504.

\bibitem[Marshall et~al.(1979)]{marshall1979inequalities}
Marshall, A.~W., Olkin, I., and Arnold, B.~C. (1979).
\emph{Inequalities: theory of majorization and its applications}.

\bibitem[Mishra et~al.(2025)]{mishra2025teacher}
Mishra, D., Sarkar, S., Sen, A., Sethuraman, J., and Yadav, S. (2025). Teacher transfers: equalizing deficits across schools.
\emph{arXiv preprint arXiv:2510.18708}.

\bibitem[Nash(1953)]{nash1953two}
Nash, J. (1953). Two-person cooperative games.
\emph{Econometrica: Journal of the Econometric Society}, pages 128--140.

\bibitem[P\'erez-Castrillo and Wettstein(2001)]{perez2001bidding}
P\'erez-Castrillo, D. and Wettstein, D. (2001). Bidding for the surplus: a non-cooperative approach to the shapley value.
\emph{Journal of economic theory}, 100(2):274--294.

\bibitem[Rothkopf et~al.(1990)]{rothkopf1990why}
Rothkopf, M.~H., Teisberg, T.~J., and Kahn, E.~P. (1990). Why are vickrey auctions rare?
\emph{Journal of Political Economy}, 98(1):94--109.

\bibitem[Rubinstein(1982)]{rubinstein1982perfect}
Rubinstein, A. (1982). Perfect equilibrium in a bargaining model.
\emph{Econometrica}, 50(1):97--109.

\bibitem[Serrano(2005)]{serrano2005fifty}
Serrano, R. (2005). Fifty years of the nash program, 1953-2003.
\emph{investigaciones econ\'omicas}, 29(2):219--258.
\end{thebibliography}
\end{document}